\documentclass[twocolumn,floatfix]{aastex62}
\usepackage{amsmath}

\usepackage{soul}

\usepackage{graphicx}
\usepackage{times}
\usepackage[normalem]{ulem}
\usepackage{color}
\usepackage{rotating}
\usepackage{enumerate}

\newcommand{\epja}{Eur.~Phys.~J.~A}
\newcommand{\prx}{PRX}
\newcommand{\rmph}{Rev. Mod. Phys.~}
\newcommand{\cqg}{{Class. Quant. Grav.}~}

\begin{document}

\title{Igniting weak interactions in neutron star postmerger accretion disks}

\correspondingauthor{Soumi De}
\email{soumide@lanl.gov}

\author[0000-0002-3316-5149]{Soumi De}
\affiliation{Computational Physics and Methods, Los Alamos National Laboratory, Los Alamos, NM 87545, USA}\affiliation{Department of Physics, Syracuse University, Syracuse, NY 13244, USA}\affiliation{Kavli Institute for Theoretical Physics, University of California, Santa Barbara, CA 93106, USA}
\author[0000-0001-6374-6465]{Daniel M.~Siegel}
\affiliation{Perimeter Institute for Theoretical Physics, Waterloo, Ontario, Canada, N2L 2Y5}
\affiliation{Department of Physics, University of Guelph, Guelph, Ontario, Canada, N1G 2W1}\affiliation{Kavli Institute for Theoretical Physics, University of California, Santa Barbara, CA 93106, USA}



\begin{abstract}
The merger of two neutron stars or of a neutron star and a black hole typically results in the formation of a post-merger accretion disk. Outflows from disks may dominate the overall ejecta from mergers and be a major source of r-process nuclei in our universe. We explore the parameter space of such disks, their outflows, and r-process yields by performing three-dimensional general-relativistic magnetohydrodynamic (GRMHD) simulations with weak interactions and approximate neutrino transport. We discuss the mapping between initial binary parameters and the parameter space of resulting disks, chiefly characterized by their initial accretion rate. We demonstrate the existence of an ignition threshold for weak interactions at around $\sim 10^{-3} M_\odot\mathrm{s}^{-1}$ for typical parameters by means of analytic calculations and numerical simulations. We find a degenerate, self-regulated, neutrino-cooled regime above the threshold and an advection dominated regime below the threshold. Excess heating in the absence of neutrino cooling below the threshold leads to $\gtrsim 60\%$ of the initial disk mass being ejected in outflows, with typical velocities $\sim (0.1-0.2)c$, compared to $\lesssim 40\%$ at $\sim (0.1-0.15)c$ above the threshold. While disks below the threshold show suppressed production of light r-process elements, disks above the threshold can produce the entire range of r-process elements in good agreement with the observed solar system abundances. Disks below the ignition threshold may produce an overabundance of actinides seen in actinide-boost stars. As gravitational-wave detectors start to sample the neutron star merger parameter space, different disk realizations may be observable via their associated kilonova emission.
\end{abstract}



\keywords{accretion, accretion disks -- magnetohydrodynamics -- gravitation -- neutrinos -- nuclear reactions: nucleosynthesis, abundances -- stars: neutron -- stars: black holes}

\section{Introduction}\label{sec:intro}

The gravitational-wave observatories advanced LIGO and Virgo, recently joined by KAGRA, are starting to make routine detections of compact binary mergers \citep{abbott_gwtc-1_2019,Abbott:2020niy,Nitz:2018imz,Nitz:2019hdf}\footnote{\href{https://gracedb.ligo.org/superevents/public/O3/}{https://gracedb.ligo.org/superevents/public/O3/}}. The first neutron-star (NS) merger GW170817 \citep{abbott_gw170817:_2017-1} and a second NS merger candidate GW190425 \citep{abbott_gw190425_2020} already probe drastically different parts of the NS binary parameter space. While the component masses of GW170817 are typical of previously known Galactic double NS systems, the total mass of GW190425 of $\sim\!3.4\,M_\odot$---if indeed a double NS system---represents an outlier by $5\sigma$ with respect to the known Galactic distribution of binary NSs \citep{abbott_gw190425_2020} (or may represent a peculiar neutron star - black hole or binary black hole system otherwise). Furthermore, while GW170817 was followed by electromagnetic counterparts across the entire electromagnetic spectrum \citep{abbott_multi-messenger_2017}, including the first unambiguous detection of a `kilonova' \citep{li_transient_1998-1,kulkarni_modeling_2005-1,metzger_electromagnetic_2010,metzger_kilonovae_2019}, GW190425 did not lead to the detection of electromagnetic counterparts, which may be due to intrinsically dim emission, the large distance of $\sim\!160$\,Mpc to the binary, and poor sky localization of $\sim\!8,300\,\mathrm{deg}^2$ \citep{abbott_gw190425_2020,antier_first_2020,lundquist_searches_2019,hosseinzadeh_follow-up_2019,coughlin_implications_2020}.

Six decades after \citet{burbidge_synthesis_1957-1} and \citet{cameron_origin_1957} realized that about half of the cosmic abundances of nuclei heavier than iron are created by rapid neutron capture onto light seed nuclei (`the r-process'), GW170817 provided the first direct observation of cosmic synthesis of such elements. The quasi-thermal emission in the ultraviolet, optical, and near-infrared was consistent with a kilonova, i.e., with being powered by the radioactive decay of r-process nuclei synthesized in the merger ejecta (see, e.g., \citealt{metzger_kilonovae_2019,siegel_gw170817_2019} for overiews discussing the interpretation of the event).

Binary neutron star (BNS) or NSBH mergers that lead to tidal disruption of the NS outside the innermost stable circular orbit can give rise to neutron-rich ejecta material conducive to the r-process in a number of ways. This includes dynamical ejecta with tidal and shock-heated components \citep{ruffert_coalescing_1997-1,rosswog_mass_1999,oechslin_relativistic_2007,hotokezaka_mass_2013,hotokezaka_progenitor_2013}, neutrino-driven and magnetically driven winds from a (meta-)stable remnant \citep{dessart_neutrino_2009,siegel_magnetically_2014,ciolfi_general_2017,ciolfi_first_2019}, and outflows from a post-merger accretion disk \citep{Metzger:2008jt,metzger_time-dependent_2008,fernandez_delayed_2013,just_comprehensive_2015,siegel_three-dimensional_2017}. The details and relative importance of these ejecta components depend on binary parameters and the unknown equation of state (EOS) of nuclear matter at supranuclear densities (see Sec.~\ref{sec:intro_ignition_threshold} for more details). In particular, the bulk of the GW170817 ejecta, specifically the material giving rise to the `red' lanthanide-bearing kilonova component, is most naturally explained by outflows from a post-merger accretion disk \citep{kasen_origin_2017,siegel_three-dimensional_2017}, while the origin of the `blue' emission in GW170817 may be due to a different source or combination of sources (\citealt{siegel_three-dimensional_2018,fernandez_long-term_2019,miller_full_2019-1,nedora_spiral-wave_2019,metzger_magnetar_2018,ciolfi_magnetically_2020}; see, e.g., \citealt{metzger_kilonovae_2019} and \citealt{siegel_gw170817_2019} for more discussion). Due to issues with chemical evolution (see \citealt{siegel_gw170817_2019} for a brief summary) and the possibility of other sources such as magneto-rotational supernovae \citep{winteler_magnetorotationally_2012,halevi_r-process_2018} and collapsars \citep{siegel_collapsars_2019} contributing significantly, it still remains an open question whether NS mergers are the dominant source of r-process elements.

Post-merger accretion disks form as a significant amount of merger debris circularized around the remnant. Such disks also provide a promising central engine to generate collimated relativistic jets needed to generate short gamma-ray bursts \citep{aloy_relativistic_2005,shibata_merger_2006,paschalidis_relativistic_2015,ruiz_binary_2016}. Numerical studies of the evolution of such accretion disks exist with various levels of approximation and computational complexity \citep{fernandez_delayed_2013,just_comprehensive_2015,siegel_three-dimensional_2017,fernandez_long-term_2019,fernandez_landscape_2020,Christie:2019lim,miller_full_2019-1,fujibayashi_properties_2017,fujibayashi_mass_2020,nedora_spiral-wave_2019,just_neutrino_2021,li_neutrino_2021}. Recent studies indicate that about 20--40\% of the disk material may be unbound into powerful neutron-rich outflows, which makes them a strong source of kilonova emission and a potentially dominant source of r-process ejecta across a wide region in NS binary parameter space (see Sec.~\ref{sec:intro_ignition_threshold} for a discussion). However, due to the computational complexity and cost, to date there exist only a few specific simulations that take all necessary physical ingredients---general relativity, magnetic fields, weak interactions, neutrino transport---into account \citep{siegel_three-dimensional_2017,siegel_three-dimensional_2018,fernandez_long-term_2019,Christie:2019lim,miller_full_2019-1,li_neutrino_2021} (see Secs.~\ref{sec:intro_ignition_threshold} and Sec.~\ref{sec:methods} for more details), and most of the post-merger parameter space and associated physics remains largely unexplored.\footnote{During the writing of this paper, \citet{fernandez_landscape_2020} published a first survey of disk models throughout the parameter space employing two-dimensional Newtonian disk simulations with an $\alpha$-viscosity to mimic angular momentum transport in the absence of magnetic fields.}

This paper presents the first exploration of the parameter space of neutrino-cooled accretion disks across two orders of magnitude in accretion rates and disk masses by means of self-consistent three-dimensional general-relativistic magnetohydrodynamic (GRMHD) simulations with weak interactions. This study is conducted in anticipation of future detections of binary mergers by LIGO, Virgo, and Kagra, which will soon sample the NS merger parameter space with many more detections. This study focuses on the transition across an ignition threshold for weak interactions, which distinguishes qualitatively distinct states and properties of such accretion disks and related parts of the neutron star binary parameter space. We begin by elaborating on this ignition threshold (Sec.~\ref{sec:intro_ignition_threshold}) and by relating post-merger disks to NS binary parameters and future detections (Sec.~\ref{sec:param_space}). A brief overview of numerical methods is provided in Sec.~\ref{sec:methods}. Section \ref{sec:results} summarizes our results, including global and local disk properties as well as r-process nucleosynthesis. Discussion and conclusions are presented in Sec.~\ref{sec:conclusion}.

\section{Physical model}

\subsection{NS post-merger disks: ignition threshold}
\label{sec:intro_ignition_threshold}

Compact accretion disks while optically thick to photons may be cooled via neutrino emission 
\citep{popham_hyperaccreting_1999,narayan_accretion_2001-1,di_matteo_neutrino_2002,beloborodov_nuclear_2003,kohri_neutrino-dominated_2005,chen_neutrino-cooled_2007,kawanaka_neutrino-cooled_2007}. At sufficiently high midplane density and temperature, weak interaction rates become high relative to the rate of radial advection of thermal energy. This gives rise to two limiting states of such disks: $(i)$ weak interactions are important and the disk is neutrino-cooled (predominantly via electron and positron capture: $e^- + p \rightarrow n + \nu_e$, $e^+ + n \rightarrow p + \bar{\nu}_e$); $(ii)$ weak interactions and neutrino cooling are negligible. In addition to changing the thermodynamics, weak interactions also change the lepton number and thus the composition of the disk and its outflows. The composition in stationary state as parametrized by the electron fraction $Y_e= n_{\rm p}/n_{\rm b}$, with $n_{\rm p}$ and $n_{\rm b}$ denoting the proton and total baryon number densities, is determined by the degree of electron/positron degeneracy \citep{beloborodov_nuclear_2003,chen_neutrino-cooled_2007,siegel_three-dimensional_2017,siegel_three-dimensional_2018}, which we explore further in this paper (Sec.~\ref{sec:disk_composition}). This has important consequences for r-process nucleosynthesis and the nature of the associated kilonova emission (Sec.~\ref{sec:nucleosynthesis}).

Weak interactions are expected to become important above a certain `ignition' threshold on the accretion rate (\citealt{Metzger:2007kj,chen_neutrino-cooled_2007,metzger_time-dependent_2008}; Sec.~\ref{sec:ignition_threshold}),
\begin{eqnarray}
    \dot{M}_{\rm ign} &\approx& \dot{\mathcal{M}}_{\rm ign}(M_{\rm BH},\chi_{\rm BH})\alpha_{\rm vis}^{\frac{5}{3}} \label{eq:Mdot_ign}\\ 
    &\approx& 2\times 10^{-3} M_\odot \,\text{s}^{-1} \left(\frac{M_{\rm BH}}{3M_\odot}\right)^{\frac{4}{3}}\left(\frac{\alpha_{\rm vis}}{0.02}\right)^{\frac{5}{3}} \label{eq:Mdot_ign_merger}.  
\end{eqnarray}
In the second step, we have evaluated the expression for the regime of post-merger disks, assuming a black-hole of mass $M_{\rm BH}=3M_\odot$ and dimensionless spin of $\chi_{\rm BH}\approx 0.8$  (see Sec.~\ref{sec:methods}) and normalizing to a dimensionless Shakura-Sunyaev viscosity coefficient $\alpha_{\rm vis} = 0.02$ (see Sec.~\ref{sec:viscosity}). While this relation has been found numerically for 1D disk solutions in Kerr spacetime \citep{chen_neutrino-cooled_2007}, we show here that the scaling $\dot{M}\propto M_{\rm BH}^{4/3} \alpha_{\rm vis}^{5/3}$ can be obtained analytically (see Sec.~\ref{sec:ignition_threshold}). Essentially, the ignition threshold can be written as a condition on the accretion rate as a result of the fact that viscous heating, neutrino cooling, and accretion rate scale with the midplane density (see Sec.~\ref{sec:ignition_threshold}).

At even higher accretion rates of $\dot{M}\gtrsim 0.1\,M_\odot \,\text{s}^{-1}$ the disk is expected to become opaque to neutrinos and at even higher rates of $\dot{M}\gtrsim 1\,M_\odot \,\text{s}^{-1}$, neutrinos start to become trapped \citep{chen_neutrino-cooled_2007}. Over a viscous timescale (Eq.~\eqref{eq:tvisc}), disks initially in such a state may evolve into an optically thin state through viscous spreading. These more extreme initial regimes are not the focus of this paper.

Many previous simulations have been performed in hydrodynamics adopting $\alpha_{\rm vis}$ as a parameter \citep{fernandez_delayed_2013,fernandez_outflows_2015,just_comprehensive_2015,fujibayashi_properties_2017,fujibayashi_mass_2020}, some of which include or model general-relativistic effects by using a pseudo-Newtonian potential. While such $\alpha$-disk models are able to qualitatively capture the evolution of disk density and angular momentum, the nature of turbulence (convection) is fundamentally different from self-consistent magnetohydrodynamic turbulence driven by the magnetorotational instability (MRI; \citealt{hawley_powerful_1992,balbus_nature_2002}). MHD disks self-consistently set an effective  $\alpha_{\rm vis}$ and thus self-consistently control the relative importance of weak interactions to viscous energy transport\footnote{This is only true in three spatial dimensions, as the anti-dynamo theorem in axisymmetry \citep{cowling_magnetic_1933} does not allow for a steady turbulent state.}; this, in turn, sets the composition of disk and outflow material and thus determines the nucleosynthetic r-process yields and kilonova colors such disks give rise to. Furthermore, while $\alpha$-disks dissipate heat generated by viscosity locally and predominantly in the disk midplane (proportional to the gas density), MHD disks dissipate a significant fraction non-locally via reconnection in low-density regions of a disk corona \citep{jiang_global_2014,siegel_three-dimensional_2018}. This difference is crucial in launching outflows, which originate from this `hot' corona with the additional help of free nuclei recombining into $\alpha$ particles \citep{siegel_three-dimensional_2018}. Indeed, a comparison between a post-merger GRMHD and $\alpha$-disk simulation shows that MHD disks are much more effective in evaporating material early on, giving rise to most of their ejecta mass during the first few hundred milliseconds, before viscous spreading (as in an $\alpha$-disk) takes over \citep{fernandez_long-term_2019}. This led to the preliminary conclusion that MHD disks can eject up to 30-40\% of their initial disk mass \citep{siegel_three-dimensional_2018,fernandez_long-term_2019}. We demonstrate in this paper (Sec.~\ref{sec:ejecta}) that the ejecta masses depend on the significance of weak interactions (cooling) and that higher relative mass loss occurs below the ignition threshold.

\subsection{NS post-merger disks: relation to binary parameters}\label{sec:param_space}

\begin{figure*}[t]
  \centering
  \includegraphics[width=16cm]{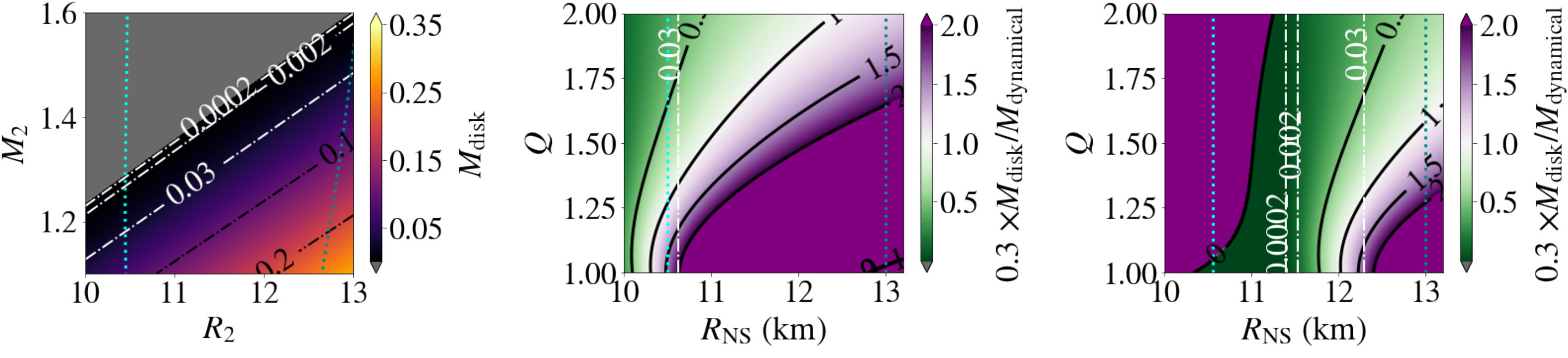}
\caption{Approximate mapping between binary parameters and accretion disk as well as ejecta masses for binary neutron star systems (based on fitting formulae to numerical relativity simulations as considered by \citealt{kruger_2020}). Left: Disk mass as a function of mass and radius of the secondary (less-massive) neutron star. Middle: Ratio of disk to dynamical ejecta mass as a function of the binary mass ratio ($Q = M_1/M_2$) and radius of the component neutron stars (assuming both stars have roughly the same radius), with the mass of the secondary fixed to $M_2 = 1.2~M_\odot$ and assuming disk ejecta to be 30\% of the initial disk mass. Right: Same as the middle panel, but for $M_2 = 1.4~M_\odot$. The dot-dashed lines (in white) delineate the disk masses $M_{\rm disk} = 0.03~M_\odot$, $0.002~M_\odot$, $0.0002~M_\odot$ considered for post-merger simulations in this work. Dotted lines (in cyan) show the lower and upper limits of neutron star radii allowed by recent observations~\citep{De:2018uhw,capano_stringent_2020,Miller:2019cac,Riley:2019yda,Landry:2020vaw}.\label{fig:bns_parameters}}
\vspace{5mm}
\end{figure*}

\begin{figure*}[t]
  \centering
  \includegraphics[width=\textwidth]{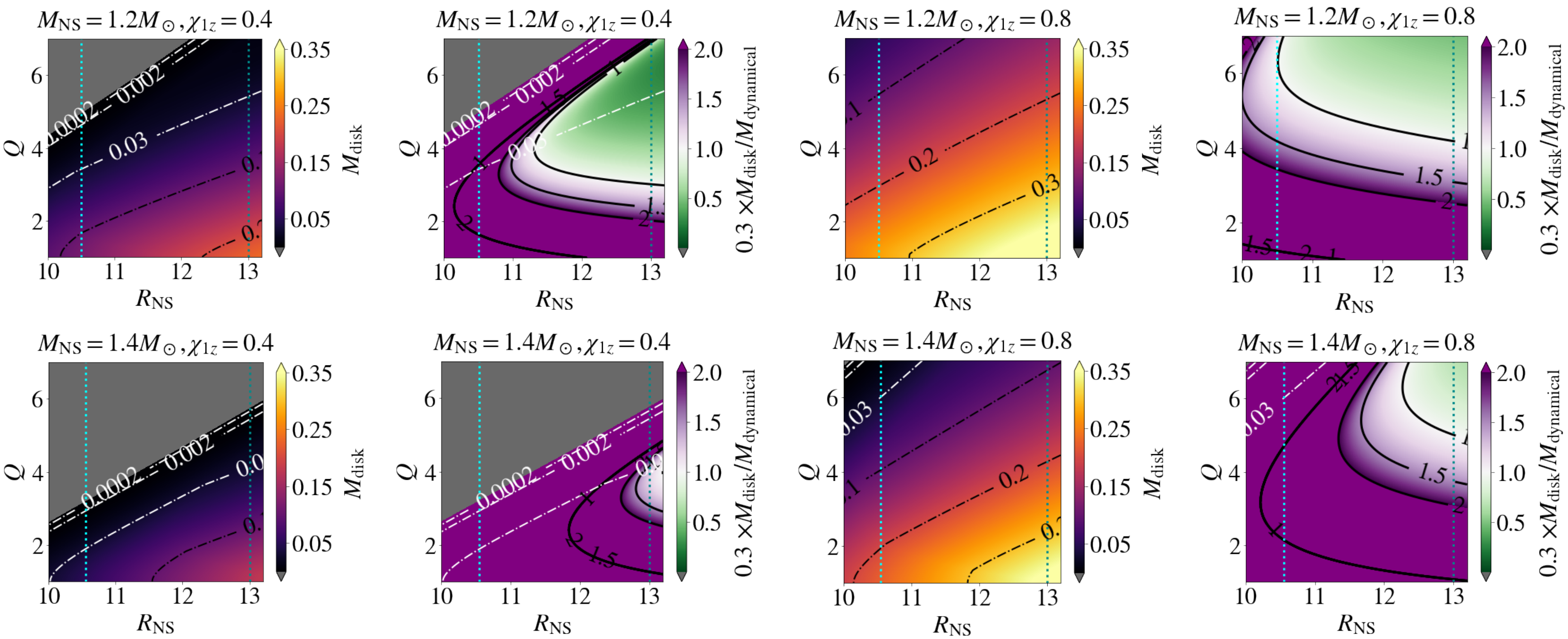}
 \caption{Approximate mapping between binary parameters and accretion disk as well as ejecta masses for neutron star--black hole binary systems (based on fitting formulae to numerical relativity simulations as considered by \citealt{foucart_remnant_2018} and \citealt{kruger_2020}). Shown are post-merger disk masses and ratios of disk to dynamical ejecta (assuming disk ejecta to be 30\% of the initial disk masses) as a function of the NS radius and the binary mass ratio $Q = M_{\rm BH}/M_{\rm NS}$ for different scenarios, including light and heavier NSs ($M_{\rm NS} = 1.2M_\odot,\, 1.4M_\odot$) and slowly or rapidly spinning BHs (dimensionless spin projected on the binary orbital angular momentum, $\chi_{1z}$, 0.4 or 0.8). The dot-dashed lines (in white) indicate the disk masses $M_{\rm disk} = 0.03~M_\odot$, $0.002~M_\odot$, $0.0002~M_\odot$ considered for the simulations in this work. Dotted lines (in cyan) show the lower and upper limits of neutron star radii allowed by observations as in Fig.~\ref{fig:bns_parameters}.\label{fig:nsbh_parameters}}
\vspace{5mm}
\end{figure*}

Figures \ref{fig:bns_parameters} and \ref{fig:nsbh_parameters} show mappings between binary parameters, accretion disk masses, and ejecta masses (dynamical and disk ejecta) for plausible binary neutron star (BNS) and neutron star black hole (NSBH) systems. Mappings from binary parameters to BNS disk masses, BNS dynamical ejecta, and NSBH dynamical ejecta are based on fitting formulae to numerical relativity simulations as considered by \citet{kruger_2020} for the respective cases. Mappings from binary parameters to NSBH disk masses are performed by calculating the mass outside the remnant black hole post merger, $M_{\rm rem}$, using the fitting formula provided by \citet{foucart_remnant_2018}, and subtracting the NSBH dynamical ejecta mass from $M_{\rm rem}$.
White lines in Figs.~\ref{fig:bns_parameters} and \ref{fig:nsbh_parameters} indicate the disk models simulated here, highlighting the regime of parameter space we focus on in this work.

For BNS systems, \citet{kruger_2020} find that disk masses extracted from existing BNS simulations can be fit to $\sim35\%$ accuracy by a formula of the type $M_{\rm disk}=M_2[aC_2 +c]^d$, which is effectively insensitive to mass ratio to leading order. Disk masses scale with the mass and compactness of the secondary (lighter) neutron star $C_2 = GM_2/(R_2c^2)$, increasing with stiffer EOSs and larger secondary component masses. For small total mass BNS systems (middle panel of Fig.~\ref{fig:bns_parameters}), mergers give rise to both dynamical ejecta and disk ejecta, irrespective of the stiffness of the EOS, while for high total mass systems (right panel of Fig.~\ref{fig:bns_parameters}) both disk masses and the amount of dynamical ejecta are reduced or even non-existent due to the fact that the BNS quickly collapse to a black hole after merger and little to no material is left outside the event horizon. 

Both the binary mass ratio and the EOS determine the dominant source of ejecta in BNS mergers. Disk ejecta dominates across most of the parameter space for small total-mass systems (middle panel of Fig.~\ref{fig:bns_parameters}), except for very small NS radii $\lesssim\!10.5$\,km and large mass ratios $Q\gtrsim 1.6$, while dynamical ejecta dominates in high total-mass systems (right panel of Fig.~\ref{fig:bns_parameters}), except for larger NS radii $\gtrsim\!12$\,km and small-to-medium mass ratios $Q\lesssim 1.4$. For low-mass BNS systems, the disks simulated here cover the parameter space for small NS radii $\lesssim\!11$\,km, while for high-mass systems they span a wide range of $11\lesssim R_{\rm NS}\lesssim 12.5$\,km. We note that the results of our highest-$M_{\rm disk}$ run can be qualitatively extrapolated to larger disk masses and thus cover most of the remaining parameter space; however, for very massive disks effects of self-irradiation of the outflows become increasingly significant, leading to a more pronounced tail of `blue' ejecta \citep{miller_full_2019-1}. Depending on the secondary mass, the disk outflows of our simulated models may or may not dominate the total ejecta of the system.

In NSBH systems, dynamical ejecta and post-merger accretion disks only arise if the NS is tidally disrupted by the BH in the binary. This disruption process requires the tidal disruption radius to reside outside of the innermost stable circular orbit of the BH, thus depending on the spin of the BH projected on the binary orbital angular momentum, $\chi_{1z}$ and its mass (and thus on the mass ratio $Q = M_{\rm BH}/M_{\rm NS}$ of the binary). Outflows from accretion disks typically dominate across most of the parameter space (cf.~Fig.~\ref{fig:nsbh_parameters}), except for light NSs ($\sim\!1.2M_\odot$) with large NS radii $\gtrsim 11.5$\,km and medium-to-large mass ratios $Q\gtrsim 3-4$, somewhat dependent on the BH spin.

In the NSBH parameter space, our models simulated here reside in the unequal mass ratio, low BH-spin regimes. 
In the regime spanned by our low-mass disk model, disk ejecta is dominant and dynamical ejecta is absent. In the regime spanned by the highest-mass disk model in this work, dynamical ejecta dominates only for light NSs ($\sim\!1.2M_\odot$) and large NS radii $\gtrsim 11.5$\,km.

\subsection{Ignition threshold and simplified 1D disk model}
\label{sec:ignition_threshold}

In this section, the basic scaling $\dot{M}_{\rm ign}\propto \alpha_{\rm vis}^{5/3}$ (Eq.~\eqref{eq:Mdot_ign}) of the accretion rate at the ignition threshold for weak interactions in an accretion disk with the dimensionless, effective Shakura-Sunyaev viscosity coefficient $\alpha_{\rm vis}$ is derived analytically. In doing so, we obtain a simplified one-dimensional accretion disk model that proves useful for further simulation analysis in Sec.~\ref{sec:results}. 

We work within a height-integrated, one-dimensional model for accretion disks in Kerr spacetime with metric $g_{\mu\nu}$ using Boyer-Lindquist coordinates $x^i = (t,r,\theta, \phi)$ (e.g., \citealt{beloborodov_super-eddington_1998,gammie_advection-dominated_1998,chen_neutrino-cooled_2007}). In such 1D models, angular averages are performed by approximating $\int {\rm d}\theta {\rm d}\phi \sqrt{g} q \simeq 4\pi H q(\theta = \pi/2)$, where $q$ represents any physical quantity, $H(r)$ is the characteristic angular half-thickness of the disk, and $g = {\rm det}(g_{\mu\nu})$. We work in the ``thin-disk'' approximation \citep{shakura_black_1973,bardeen_rotating_1972,novikov_astrophysics_1973}, well justified if neutrino cooling is significant, in which terms in $(H/r)^2$ are neglected and gas pressure is negligible. This results in the assumption that the fluid orbits with four-velocity $u^\mu = (u^t,u^r, 0, u^\phi)$ and Keplerian angular velocity $\Omega \equiv u^\phi/u^t = c/(r\sqrt{2r/r_{\rm g}} + \chi_{\rm BH} r_{\rm g}/2)$. Here, $r_{\rm g}=2GM_{\rm BH}/c^2$ is the gravitational radius, and $M_{\rm BH}$ and $\chi_{\rm BH}$ denote, as before, the mass and the dimensionless spin of the black hole, respectively. In this thin-disk limit, the equation of vertical hydrodynamical equilibrium for the disk can be written as (cf.~\citealt{abramowicz_accretion_1997,beloborodov_super-eddington_1998})
\begin{equation}
	\left(\frac{H}{r}\right)^2 = \frac{2 r}{c^2 r_{\rm g} J(\chi_{\rm BH},r)}\frac{p}{\rho},\mskip40mu 
\end{equation}
or
\begin{equation}
	c_{\rm s} = c\left(\frac{J(\chi_{\rm BH},r) r_{\rm g}}{2r}\right)^{\frac{1}{2}}\frac{H}{r}, \label{eq:vertical_equilibrium}
\end{equation}
where $c_{\rm s}=\sqrt{p/\rho}$ is the isothermal sound speed, and
\begin{equation}
	J(\chi_{\rm BH},r) \equiv \frac{2\left(r^2 - \chi_{\rm BH} r_{\rm g}\sqrt{2r_{\rm g}r}+ 3\chi_{\rm BH}^2 r_{\rm g}^2 / 4 \right)}{2r^2 - 3r_{\rm g}r + \chi_{\rm BH} r_{\rm g}\sqrt{2r_{\rm g}r}}.
\end{equation}
The equation of baryon number conservation reads
\begin{equation}
	\dot{M} = -2\pi r c u^r \Sigma, \label{eq:baryon_conservation}
\end{equation}
where $\Sigma = 2H\rho$ is the disk surface density. From the equations of energy and angular momentum conservation, one can derive the identity \citep{page_disk-accretion_1974,chen_neutrino-cooled_2007}
\begin{equation}
	2\nu \Sigma r \sigma^{r}_\phi = -\frac{c r_{\rm g}}{4\pi} \dot{M} F(x,\chi_{\rm BH}), \label{eq:shear_identity}
\end{equation}
where $\nu$ is the kinematic viscosity, 
\begin{equation}
	\sigma^r_\phi = (1/2) c g^{rr} g_{\phi\phi}\sqrt{-g^{tt}}\gamma^3 ({\rm d}\Omega/{\rm d}r)
\end{equation}	
is the shear, with $\gamma=u^t/\sqrt{-g^{tt}}$ being the Lorentz factor of the fluid measured by a zero-angular momentum observer, and $F(x,\chi_{\rm BH})$ is a function of dimensionless radius $x\equiv \sqrt{2r/r_{\rm g}}$ and black-hole spin given in Appendix \ref{app:ignition_threshold} for completeness.

Substituting Eq.~\eqref{eq:shear_identity} into Eq.~\eqref{eq:baryon_conservation} and using Eq.~\eqref{eq:vertical_equilibrium} one finds
\begin{equation}
	u^r = \frac{8}{3} \left(\frac{J(r,\chi_{\rm BH}) r}{2c^2 r_{\rm g}}\right)^{\frac{1}{2}} \frac{\sigma^r_\phi}{F(x,\chi_{\rm BH})}\alpha_{\rm vis} \left(\frac{H}{r}\right)^2, 
\end{equation}
where we have adopted the Shakura-Sunyaev parametrization $\nu = \frac{2}{3}\alpha_{\rm vis} c_{\rm s} H$ with viscosity coefficient $\alpha_{\rm vis}$. Substituting this back into Eq.~\eqref{eq:baryon_conservation}, we obtain
\begin{eqnarray}
	\dot{M} &=& - \frac{32\pi r_{\rm g}^2}{3\sqrt{2}} J^{\frac{1}{2}}(r,\chi_{\rm BH}) \mskip-5mu\left(\frac{r}{r_{\rm g}}\right)^{\frac{5}{2}} \mskip-15mu\frac{\sigma^r_\phi}{F(x,\chi_{\rm BH})}\alpha_{\rm vis} \rho \mskip-5mu\left(\frac{H}{r}\right)^3  \label{eq:Mdot_1D_1} \\
		&=& 2\sqrt{2}\pi c r_{\rm g}^2 S^{-1} J^{\frac{1}{2}} \left(\frac{r}{r_{\rm g}}\right)^{\frac{3}{2}} \alpha_{\rm vis} \rho \left(\frac{H}{r}\right)^3,  \label{eq:Mdot_1D_2}
\end{eqnarray}
where we have introduced the function $S(r,\chi_{\rm BH})\equiv-(3/8)c(r_{\rm g}/r) [F(x,
\chi_{\rm BH})/\sigma^r_\phi]$, which varies between zero at the marginally stable orbit and unity at $r\rightarrow\infty$.

The viscous heating rate per unit area of the disk is given by $Q^+ = 2\nu\Sigma h c \sigma^r_\phi \sqrt{-g^{tt}}\gamma ({\rm d}\Omega/{\rm d}r)$, where $h\approx 1$ is the specific enthalpy of the fluid in the thin disk \citep{beloborodov_super-eddington_1998}. Using the identity \eqref{eq:shear_identity} one can rewrite this as
\begin{equation}
	Q^+ = \frac{c^2}{4\pi}\left(\frac{r}{r_{\rm g}}\right)^{-1} F(x,\chi_{\rm BH}) \mathcal{Q}(r,M_{\rm BH},\chi_{\rm BH}) \dot{M}, \label{eq:Qplus}
\end{equation}
where $\mathcal{Q}(r,M_{\rm BH},\chi_{\rm BH})\equiv -u^t ({\rm d}\Omega/{\rm d}r)$. 

We assume that cooling of the disk is dominated by electron and positron capture (URCA cooling). Ignoring final state blocking in the neutrino phase space, the cooling rate per unit area of the disk is then approximately given by \citep{tubbs_neutrino_1975,bruenn_stellar_1985,qian_nucleosynthesis_1996,popham_hyperaccreting_1999}
\begin{equation}
	Q^-_\nu = 2H \mathcal{C}_\nu \rho T^6. \label{eq:Qminus_nu_def}
\end{equation}
Here, $\mathcal{C}_\nu$ is a constant times the mass fraction of nucleons $X_{\rm nuc}$, which is roughly unity in the inner parts of the accretion disk where photodisintegration breaks down nuclei into neutrons and protons once $T\sim 10^{10}$\,K. At sufficiently small $\dot{M}$ (low midplane density), the pressure is dominated by radiation pressure, $p = \frac{11}{12}a_{\rm SB} T^4$, where contributions of relativistic electron-positron pairs have been included \citep{popham_hyperaccreting_1999}. Substituting into Eq.~\eqref{eq:vertical_equilibrium}, this yields the disk midplane temperature
\begin{equation}
	T = \left(\frac{6}{11}\frac{c^2}{a_{\rm SB}}\right)^{\frac{1}{4}} J^{\frac{1}{4}}(r,\chi_{\rm BH}) \left(\frac{r}{r_{\rm g}}\right)^{-\frac{1}{4}} \left(\frac{H}{r}\right)^{\frac{1}{2}} \rho^{\frac{1}{4}}. \label{eq:T_1D}
\end{equation}
Using this relation in Eq.~\eqref{eq:Qminus_nu_def} together with Eq.~\eqref{eq:Mdot_1D_2}, one obtains the following expression for neutrino cooling:
\begin{eqnarray}
	Q^-_\nu &=& 2 \mathcal{C}_\nu \left(\frac{1}{2\sqrt{2}\pi c}\right)^{\frac{5}{2}} \left(\frac{6}{11}\frac{c^2}{a_{\rm SB}}\right)^{\frac{3}{2}} \\
	&&\times r_{\rm g}^{-4} S^{\frac{5}{2}} J^{\frac{1}{4}} \left(\frac{r}{r_{\rm g}}\right)^{-\frac{17}{4}}\left(\frac{H}{r}\right)^{-\frac{7}{2}} \alpha_{\rm vis}^{-\frac{5}{2}}\dot{M}^{\frac{5}{2}}. \label{eq:Qminus_nu}
\end{eqnarray}

Adopting the condition $Q^-_\nu/Q^+ = 1/2$ for weak interactions to become energetically significant, one can employ the expressions \eqref{eq:Qplus} and \eqref{eq:Qminus_nu} to formulate this as a condition on the accretion rate:
\begin{eqnarray}
	\dot{M}_{\rm ign} &=& 
	\frac{11}{12}\sqrt{2}^{\frac{5}{3}}\pi a_{\rm SB} \mathcal{C}_\nu^{-\frac{2}{3}}\\
	&&\times r_{\rm g}^{\frac{8}{3}} S^{-\frac{5}{3}} J^{-\frac{1}{6}}
	F^{\frac{2}{3}} \mathcal{Q}^{\frac{2}{3}}
	\mskip-5mu\left(\frac{r}{r_{\rm g}}\right)^{\frac{13}{6}}\mskip-5mu\left(\frac{H}{r}\right)^{\frac{7}{3}} \mskip-5mu\alpha_{\rm vis}^{\frac{5}{3}} \\
		&\equiv& \dot{\mathcal{M}}_{\rm ign}(r,M_{\rm BH},\chi_{\rm BH}) \alpha_{\rm vis}^{\frac{5}{3}}.
\end{eqnarray}
where $\dot{\mathcal{M}}_{\rm ign}(r,M_{\rm BH},\chi_{\rm BH})$ is a function that depends on the black-hole parameters. Apart from $S(r,\chi_{\rm BH})$, which must be calculated numerically (but may be approximated analytically), $\dot{\mathcal{M}}_{\rm ign}(r,M_{\rm BH},\chi_{\rm BH})$ can be analytically evaluated on a horizon-scale $r\sim r_{\rm g}$ to provide the characteristic accretion rate onto the black hole. Noting that at $r\sim r_{\rm g}$, $\mathcal{Q}\propto r_{\rm g}^{-2}$, we find that the critical accretion rate onto the black hole scales as
\begin{equation}
    \dot{M}_{\rm ign} \propto M_{\rm BH}^{\frac{4}{3}} \alpha_{\rm vis}^{\frac{5}{3}},
\end{equation}
with a prefactor of order unity that depends on the black hole spin $\chi_{\rm BH}$.

\section{Numerical Methods}
\label{sec:methods}

\subsection{Simulation setup}
\label{subsec:sim_setup}

We perform simulations in ideal GRMHD and full 3D with a fixed background spacetime for computational efficiency using the code and numerical setup described in \citet{siegel_three-dimensional_2018}. The code is based on \texttt{GRHydro}~\citep{mosta_grhydro:_2014} and makes use of the \texttt{Einstein Toolkit}\footnote{\href{http://einsteintoolkit.org}{http://einsteintoolkit.org}} \citep{maria_babiuc-hamilton_einstein_2019,loffler_einstein_2012,Schnetter:2003rb,Goodale:2002a,Thornburg:2003sf}, with neutrino interactions implemented via a leakage scheme based on \citet{bruenn_stellar_1985} and \citet{ruffert_coalescing_1996}, and follows the implementation of \citet{galeazzi_implementation_2013} and \citet{radice_dynamical_2016}. Thermodynamic properties of matter are based on the Helmholtz EOS \citep{timmes_accuracy_1999,timmes_accuracy_2000}, and we compute abundances of nuclei at a given density, temperature, and electron fraction $Y_e$, assuming nuclear statistical equilibrium.

\begin{figure*}[t]
  \includegraphics[width=\textwidth]{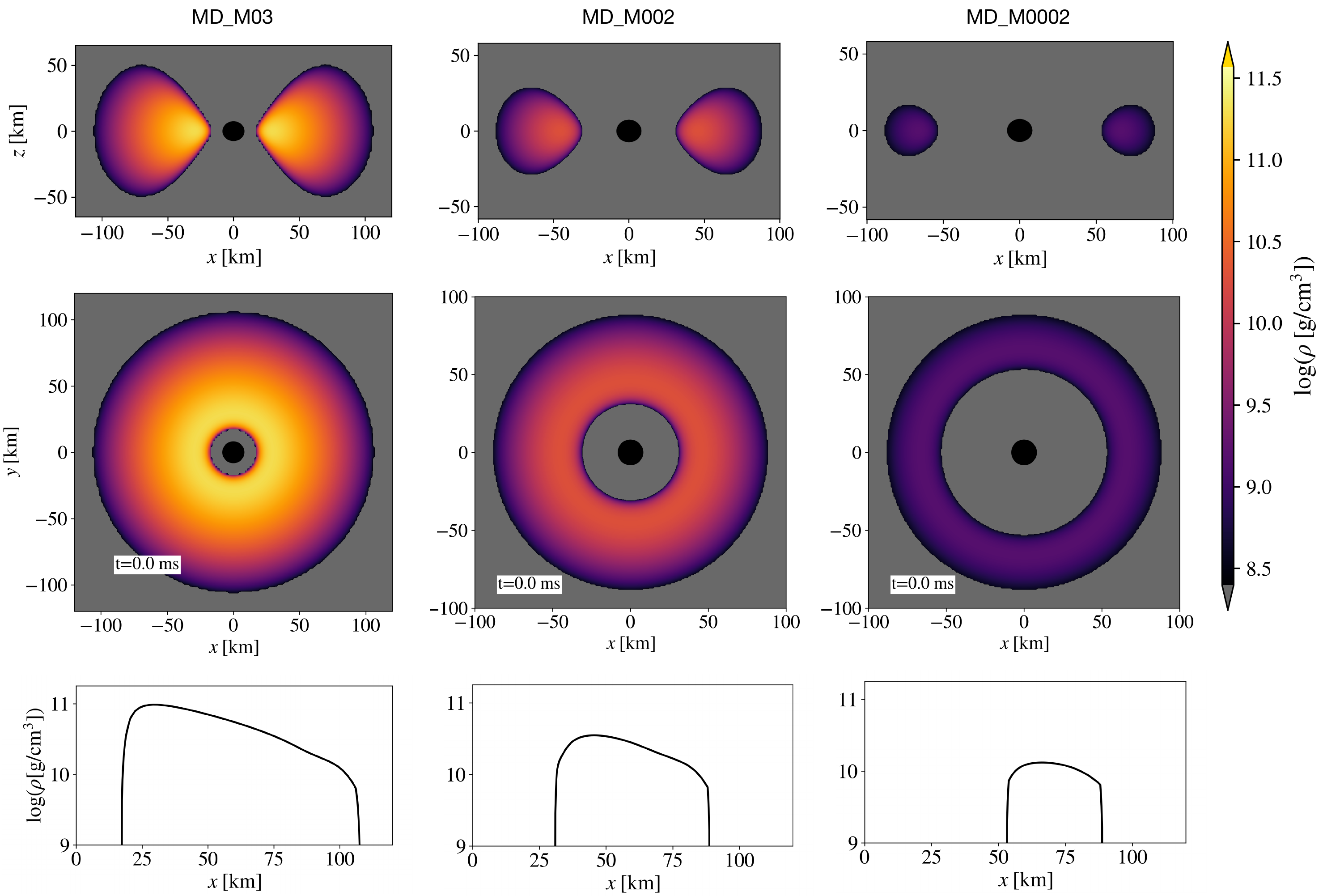}
 \caption{Initial disks in the three simulations performed in this work. The top and middle panels show density slices along the equatorial and meridional plane, respectively, at $t = 0$~ms. The bottom panels show radial profiles of density. 
 \label{fig:ini_torus}}
\vspace{5mm}
\end{figure*}

The simulations include a Kerr black hole of mass $3.0 M_\odot$ and dimensionless spin $\chi_{\rm BH} = 0.8$, initially surrounded by a torus of constant specific angular momentum, small constant specific entropy of $8~k_{\rm B}$ per baryon, and initial electron fraction $Y_{\rm e} = 0.1$; we refer to Tab.~\ref{tab:ini_config_torus} for a summary of initial torus properties of the simulation runs. Figure \ref{fig:ini_torus} shows disk densities for the initial configurations.
Run \texttt{MD\_M03} has been discussed before \citep{siegel_three-dimensional_2017,siegel_three-dimensional_2018}, and is further elaborated on here by comparing it to the two new runs \texttt{MD\_M002} and \texttt{MD\_M0002}, which represent lighter accretion disks. The BH-disk problem is formulated in Cartesian, horizon-penetrating Kerr-Schild coordinates. The black-hole mass and spin reflect typical NS merger scenarios (see Sec.~\ref{sec:param_space}). The black hole spins in the case of prompt black hole formation from BNS mergers are typically not larger than $\chi_{\rm BH} \approx 0.8$~\citep{kiuchi_longterm_2009,rezzolla_accurate_2010,bernuzzi_mergers_2014,kastaun_black_2013}, and black hole spins in case of delayed black hole formation are $\chi_{\rm BH} \lesssim 0.7$ \citep{sekiguchi_dynamical_2016}. Furthermore, $\chi_{\rm BH} \sim 0.8$ is significant enough to disrupt the NS and lead to a post-merger accretion disk in NSBH mergers across a wide range in mass ratio (\citealt{foucart_black-hole-neutron-star_2012}; see the discussion in Sec.~\ref{sec:param_space}). Our initial tori masses are chosen to reflect a mass range covering the ignition threshold for weak interactions in post-merger disks (cf.~Sec.~\ref{sec:intro_ignition_threshold}) and is typical both for BNS and NSBH scenarios (Sec.~\ref{sec:param_space}).

\begin{table*}
\caption{Initial configurations of the accretion disks before relaxation. From left to right: black-hole mass and dimensionless spin, disk mass, inner and outer radius of the disk, radius at maximum density, specific entropy, electron fraction, and maximum magnetic-to-fluid pressure ratio.}
\begin{centering}
\begin{tabular}{cccccccccc}
\hline
Run  & $M_{\rm BH}$ & $\chi_{\rm BH}$ & $M_{\rm d,0}$ & $R_{\rm in,0}$ & $R_{\rm out,0}$  & $R_0$ & $s_0$ & $Y_{\rm e,0}$ & $p_b/p_{\rm f}$\\
& [$M_{\odot}$] & & [$M_{\odot}$] & [km] & [km] &  [km] & [$k_B$/b] & & \\
\hline 
MD\_03 & 3 & 0.8 & 0.03 & 17.71 & 106.27 & 30 & 8 & 0.1 & $< 5\times 10^{-3}$ \\
MD\_002 & 3 & 0.8 & 0.002 & 31 & 88.56 & 45.61 & 8 & 0.1 & $< 5\times 10^{-3}$ \\
MD\_0002 & 3 & 0.8 & 0.0002 & 53.14 & 88.56 & 66.27 & 8 & 0.1 & $< 5\times 10^{-3}$ \\
\hline 
\end{tabular}
\par\end{centering}
\label{tab:ini_config_torus}
\end{table*}

The tori are initialized with weak poloidal magnetic seed fields, confined to the interior of the tori and defined by the magnetic vector potential $A^r = A^{\theta} = 0$ and $A^{\phi} = A_b$ max$\{p - p_{\rm cut}, 0\}$. Here, $p$ denotes the fluid pressure, $p_{\rm cut}$ is the pressure below which the magnetic field is set to zero, and $A_b$ sets the initial field strength. Here, $p_{\rm cut} \approx 1\times 10^{-2} p_{\rm max}$ in all cases, where $p_{\rm max}$ is the pressure at maximum density in the torus. We choose $p_{\rm cut}$ such that the magnetic field covers the bulk volume of the torus, while preventing it from becoming buoyant in the outermost layers and violently breaking out of the torus at the start of the simulation. We adjust $A_b$ such that the magnetic-to-fluid pressure ratio in the torus is a small value, $p_{\rm B}/p_{\rm f} < 5\times 10^{-3}$. This ratio provides relatively `weak' (dynamically unimportant) initial magnetic field strengths with maximum values of $\approx\!3\times 10^{14}$ G for \texttt{MD\_M03}, $\approx\!6.4\times 10^{13}$ G for \texttt{MD\_M002}, and $\approx\!1.3\times 10^{13}$ G for \texttt{MD\_M0002}.

The initial torus is embedded in a tenuous atmosphere with $T = 10^5$ K, $Y_{\rm e} = 1$, and $\rho \approx$ 37 g cm$^{-3}$, $\rho \approx$ 3.7 g cm$^{-3}$, and $\rho \approx$ 0.37 g cm$^{-3}$ for runs \texttt{MD\_M03}, \texttt{MD\_M002}, and \texttt{MD\_M0002}, respectively. The density and temperature are chosen such that they are sufficiently low to neither impact the dynamics nor the composition of the disk outflows. The atmosphere densities are set to approximately scale with the maximum density of the accretion disk during the evolution (cf.~Tab.~\ref{tab:torusenergytab}). The total atmosphere mass of the entire computational domain is $3.8\times 10^{-5} M_\odot$ for \texttt{MD\_M03}, $3\times 10^{-7} M_\odot$ for \texttt{MD\_M002}, and $3\times 10^{-8}$ for \texttt{MD\_M0002}, orders of magnitude smaller than the disk ejecta (cf.~Tab.~\ref{tab:torusenergytab}); over a volume of radius 1000\,km, which we consider as the minimum radius for outflow material to be unbound from the BH-disk system, the corresponding atmosphere masses are $1.8\times 10^{-8} M_\odot$ for \texttt{MD\_M03}, $1.8\times 10^{-9} M_\odot$ for \texttt{MD\_M002}, and $1.8\times 10^{-8} M_\odot$ for \texttt{MD\_M0002}. At the chosen atmosphere temperature of $T = 10^5$\,K weak interations are frozen out.

The computational domain represents a Cartesian grid hierarchy centered around the black hole. For \texttt{MD\_M03}, the grid has eight refinement levels, with an extent in each coordinate direction of $1.53\times 10^4$ km. For \texttt{MD\_M002} and \texttt{MD\_M0002}, the grid has seven refinement levels, with an extent in each coordinate direction of $1.14\times 10^4$ km. The initial tori have diameters of 240~km, 206~km, and 206~km for simulations \texttt{MD\_M03}, \texttt{MD\_M002}, and \texttt{MD\_M0002}, respectively. The initial tori are encompassed by the finest refinement level of the corresponding grid hierarchy. Following previous work \citep{siegel_magnetorotational_2013,siegel_three-dimensional_2018,siegel_collapsars_2019}, the finest resolution is $\Delta_{xyz} \approx 850$\,m for all simulations, chosen such that the MRI is well resolved in the stationary turbulent state of the disk (typically by at least ten grid points per fastest-growing MRI mode), which ensures convergence of global observables (see, e.g., \citealt{siegel_collapsars_2019}).

\begin{figure*}[t]
  \includegraphics[width=\textwidth]{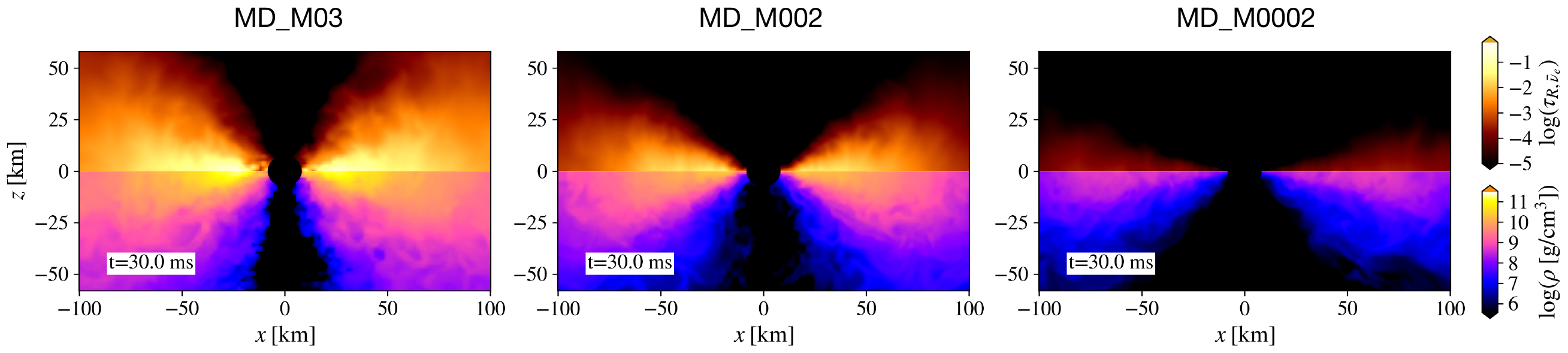}
 \caption{Slices of rest-mass density and optical depth for $\bar{\nu}_e$ neutrinos in the meridional plane at 30\,ms shortly after the disks have relaxed into a quasi-stationary state. This state represents the actual initial data for our simulations. All disks are optically thin to neutrinos. \label{fig:optical_depth}}
\vspace{5mm}
\end{figure*}

Angular momentum transport is mediated by magnetic turbulence, driven by the MRI, in our setups. The initial tori undergo a relaxation phase with self-consistent magnetic-field amplification, and settle into a quasi-stationary phase at $\sim\!20$ ms. We consider the relaxed state of the disks at this time as the actual initial data for our simulations, and exclude the early transient phase from our analysis---in particular, all material accreted onto the black hole or ejected via outflows during this phase. Figure \ref{fig:optical_depth} shows snapshots of the disks just after having settled into the quasi-stationary phase, indicating that all configurations are optically thin to neutrinos.

\subsection{Diagnostics}\label{subsec:diagnostics}

In order to monitor certain physical quantities in the disk, we compute radial profiles of a quantity $\chi(\varpi)$ by performing azimuthal, density-weighted averages, integrating up to one scale height of the disk:

\begin{equation}\label{eq:radial_av}
    \langle \chi \rangle_{{\rm azim}, z_H} = \frac{\int^{z_H}_{-z_H} \int_0^{2\pi} \chi \hat D \varpi d\phi dz}{\int^{z_H}_{-z_H} \int_0^{2\pi} \hat D \varpi d\phi dz}.
\end{equation}
Here, $\varpi = \sqrt{x^2 + y^2}$ is the cylindrical radius, $\hat D = \sqrt{\gamma} \rho W$ is the conserved rest-mass density as seen by the Eulerian observer moving normal to the spatial hypersurfaces in the 3+1 split of Kerr-Schild spacetime, $\gamma$ is the determinant of the spatial metric $\gamma_{ij}$ in 3+1 split, and $W$ the Lorentz factor of the fluid. The density scale height $z_H$ is defined as
\begin{equation}
    z_H (\varpi) = \frac{\int \int_0^{2\pi} |z| \hat D \varpi d\phi dz}{\int \int_0^{2\pi}\hat D \varpi d\phi dz}.
\end{equation}
By integrating up to the local density scale height, we exclude the disk corona and winds from the calculation. For some quantities, temporal averages $\langle\cdot\rangle_t$ (cf., e.g., Eqs.~\eqref{eq:alpha_visc_def_1D} and \eqref{eq:alpha_visc_def_0D}) over a specified time window are taken, in order to reduce the effect of temporal fluctuations of a turbulent medium.

For some disk quantities, it is useful to compute the rest-mass density average and study its evolution over time. We define this averaging as
\begin{equation}\label{eq:rest_mass_dens}
 \langle \chi \rangle_{\hat D} = \frac{\int \chi \hat D d^3x}{\int \hat D d^3x}.
\end{equation}
In some cases, this average is calculated by integrating only up to the local scale height,
\begin{equation}\label{eq:rest_mass_dens_scaleht}
    \langle \chi \rangle_{\hat D, z_H} = \frac{\int^{z_H}_{-z_H} \chi \hat D \varpi dz}{\int^{z_H}_{-z_H} \hat D \varpi dz},
\end{equation}
which allows us to explicitly exclude the disk corona and disk wind regions.

\section{Numerical Results}
\label{sec:results}

We proceed by first discussing our results for global disk properties, such as MHD mediated angular momentum transport (Sec.~\ref{sec:viscosity}), accretion (Sec.~\ref{sec:accretion}), neutrino emission (Sec.~\ref{sec:neutrino_emission}), disk ejecta (Sec.~\ref{sec:ejecta}), before discussing disk evolution locally in terms of compositional changes and weak interactions (Sec.~\ref{sec:disk_composition}), and nucleosynthesis from the disk outflows (Sec.~\ref{sec:nucleosynthesis}).

\subsection{Global properties}
\label{sec:global_properties}

\begin{table*}
\caption{Various properties of the accretion disks simulated here. From left to right: i) average density of the inner accretion disk, ii) disk mass after relaxation, iii) accretion rate after relaxation, iv) estimated total unbound disk outflow mass ($M_{\rm ej}$; `ejecta') in units of the effective initial disk mass after relaxation, v) with mean electron fraction, vi) maximum total neutrino luminosity in electron and anti-electron neutrinos, vii) effective $\alpha$-viscosity parameter of the inner accretion disk, viii) viscous timescale and ix) total simulated time for the respective runs. The averages $\bar{\rho}_{\rm d}$ and $\bar \alpha_\mathrm{vis}$ with the quoted uncertainties represent density-weighted average, maximum, and minimum values around a 10\,ms window centered at $t = 30$\,ms (see the text for details). The values for $\dot{M}_{d}$ with quoted uncertainties represent the average, maximum, and minimum values over the same time window at $t = 30$\,ms. The values for $L_{\nu,\mathrm{max}}$ represent the average value extracted from the neutrino luminosity over a 10\,ms window centered at the global peak luminosity of anti-electron neutrinos.
}
\begin{centering}
\begin{tabular}{ccccccccccc}
\hline
Run  & $\bar{\rho}_{\rm d}$ & $M_{\rm d,20}$ & $\dot{M}_{d}$  & $M_{\rm ej}$ & $\bar{Y}_e$ & $L_{\nu,\mathrm{max}}$ & $\bar \alpha_\mathrm{vis}$ & $t_{\rm vis}$ & $t_{\rm sim}$\\
 & [$\mathrm{g}\,\mathrm{cm}^{-3}$] & [$10^{-2}M_{\odot}$] & [$M_{\odot}\,\mathrm{s}^{-1}$] &  [$M_{\rm d,20}$] & & [$\mathrm{erg}\,\mathrm{s}^{-1}$] & $\times 10^{-2}$ & [ms] & [ms]\\
\hline 
MD\_03 & $5.53_{-1.4}^{+2.2}\times 10^{10}$ & 1.89 & $2.6_{-0.7}^{+1.4}\times 10^{-1}$ & 0.17 & 0.174 & 2.1$\times 10^{52}$ & $1.76_{-1.7}^{+2.8}$ & 279 & 381\\
MD\_002 & $6.6^{+0.5}_{-0.9}\times 10^9$ & 0.184 & $1.14_{-0.4}^{+0.4}\times 10^{-2}$ & 0.22 & 0.114 & 14$\times 10^{50}$ & $1.01_{-0.6}^{+1.5}$ & 874 & 309\\
MD\_0002 & $3.4^{+0.7}_{-0.6}\times 10^8$ & 0.0199 & $3.79_{-1.7}^{+2.5}\times 10^{-4}$ & 0.30 & 0.101 & 2.1$\times 10^{48}$ & $1.14_{-1.1}^{+2.0}$ & 1731 & 294\\

\hline 
\end{tabular}
\par\end{centering}
\label{tab:torusenergytab}
\end{table*}

\subsubsection{MHD turbulence \& effective viscosity}
\label{sec:viscosity}

\begin{figure*}[t]
  \includegraphics[width=\textwidth]{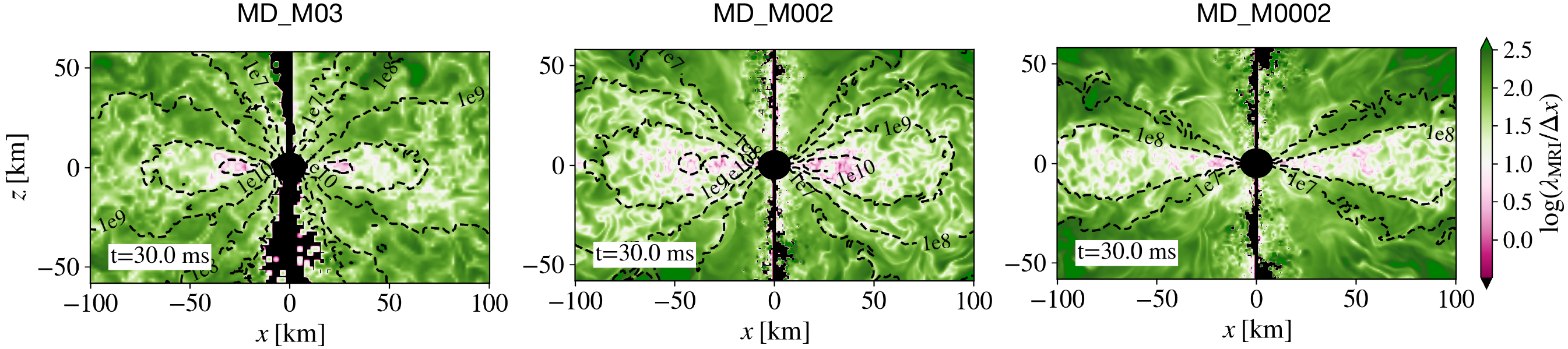}
 \caption{Meridional slices through the simulated disks, showing the number of grid points per fastest growing MRI wavelength at $t = 30$~ms. Overlaid are contours of density at $\rho = 10^7, 10^8, 10^9,$ and $10^{10}$ g cm$^{-3}$.\label{fig:lambda-mri}}
\vspace{5mm}
\end{figure*}

Soon after the start of the simulations, our accretion disks show vigorous magnetic turbulence, triggered by the MRI, a local fluid instability developed in differentially rotating magnetized fluids~\citep{velikhov_notitle_1959,chandrasekhar_stability_1960,balbus_powerful_1991,balbus_instability_1998,balbus_enhanced_2003,armitage_dynamics_2011}. The initial weak magnetic field in the simulations, which is dynamically not important and purely poloidal, is amplified by the MRI at an exponential rate. A toroidal field component is initially generated and amplified by magnetic winding, before the toroidal field becomes susceptible to the MRI as well. This combination of the MRI and magnetic winding causes an overall increase of the maximum total magnetic field strength by one to two orders of magnitude from its initial value to saturation. At $t\approx 20$~ms, a steady turbulent state is achieved by the disk and the magnetic field, self-consistently amplified by the MRI, reaches saturation, losing memory of the initial magnetic field configuration. We refer to \citet{siegel_three-dimensional_2018} for a more detailed discussion on how this turbulent state arises. The resolution employed here (see Sec.~\ref{subsec:sim_setup}) guarantees a converged saturation level for the magnetic field strength of each run; thus, the efficiency of MHD mediated angular momentum transport does not depend on the particular resolution employed here \citep{siegel_collapsars_2019}. Figure \ref{fig:lambda-mri} shows the number of grid points per fastest growing MRI wavelength, $\lambda_{\mathrm{MRI}}/\Delta{x}$, in the three accretion disk models, shortly after the disks have reached a quasi-stationary state. The MRI can be considered to be well resolved when $\lambda_{\mathrm{MRI}}/\Delta{x} \gtrsim 10$. Figure \ref{fig:lambda-mri} shows that at $t = 30$~ms, $\lambda_{\mathrm{MRI}}/\Delta{x}$ is at least $\sim\!10$ in all the models, and it continues to be so through the rest of the simulations. This condition can be violated locally in space and time due to the turbulent nature of the accretion flow.

MHD turbulence in the disk operates as large-scale viscosity, which can be parametrized by an effective Shakura-Sunyaev viscosity $\alpha_{\rm vis}$. We define this quantity as the ratio of total stress to fluid pressure,
\begin{equation}
    \alpha_{\rm vis} (\varpi) = \frac{\langle \langle |T^{r,\phi}|\rangle_{{\rm azim}, z_H}\rangle_t}{\langle \langle p \rangle_{{\rm azim}, z_H}\rangle_t}, \label{eq:alpha_visc_def_0D}
\end{equation}
where $T^{r,\phi}$ is the $r-\phi$ component of the stress-energy tensor in the frame comoving with the fluid, and $p$ the fluid pressure. Here, time averages are taken over a few neighboring data snapshots.

\begin{figure}[t]
  \includegraphics[width=\columnwidth]{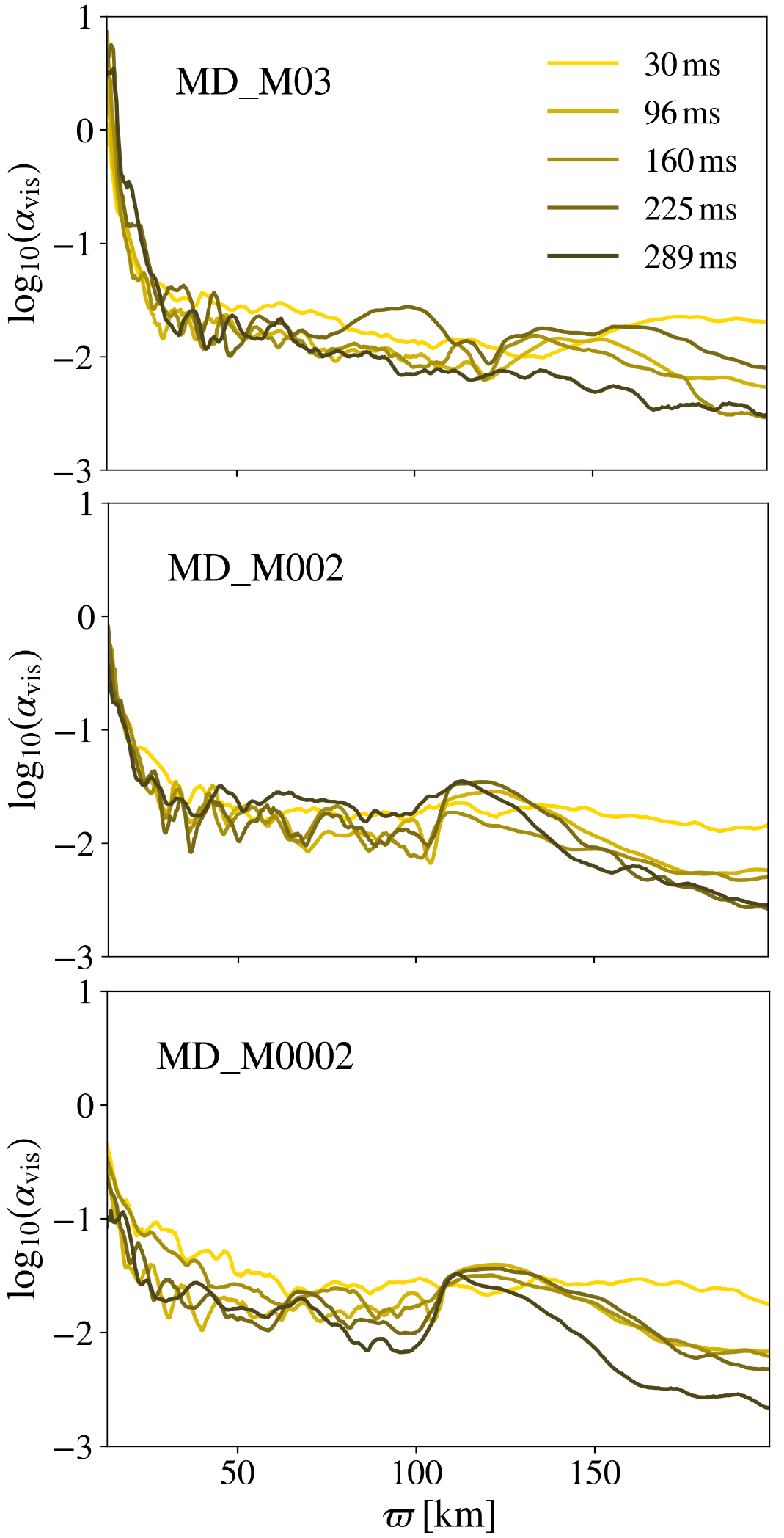}
 \caption{Radial profiles of the effective Shakura-Sunyaev viscosity parameter $\alpha_{\rm vis}$ defined in Eq.~\eqref{eq:alpha_visc_def_0D} at different times during the evolution for the three simulation runs in this work. \label{fig:alpha-viscosity}}
\vspace{5mm}
\end{figure}

Figure \ref{fig:alpha-viscosity} shows radial profiles of $\alpha_{\rm vis}$, computed using Eq.~\eqref{eq:radial_av}, at different times during the disks' evolution, for all three simulation runs. We find that $\alpha_{\rm vis}$ is roughly constant among the simulations, independent of the initial disk mass. Table \ref{tab:torusenergytab} reports global values for $\alpha_{\rm vis}$ for each simulation. The latter values describe the accretion state of the inner disk. We compute these values as an average over a 10~ms time-window centered at $t = 30$~ms, extracted from the time evolution of the absolute value of the rest-mass density average of $\alpha_{\rm vis}$,
\begin{equation}
    \alpha_{\rm vis} = \frac{\langle \langle |T^{r,\phi}|\rangle_{\hat{D}, z_H}\rangle_t}{\langle \langle p \rangle_{\hat{D}, z_H}\rangle_t}. \label{eq:alpha_visc_def_1D}
\end{equation}
The rest-mass density average is calculated following the procedure in Equation \eqref{eq:rest_mass_dens_scaleht}, restricting to the region between 30-175\,km from the centers of the black holes. By applying this restriction along the radial extent, only the inner disk regions that can be directly related to the accretion rate onto the black hole are used in the calculation; the data of the outer parts of the disks that start to viscously spread are excluded from the calculation. Also excluded are the innermost regions that show a rise of $\alpha_\mathrm{visc}(\varpi)$ toward the innermost stable circular orbit and the black-hole horizon due to increasing mean magnetic field strengths (\citealt{penna_shakura-sunyaev_2013}; cf.~Fig.~\ref{fig:alpha-viscosity}). As indicated by the radial profiles in Fig.~\ref{fig:alpha-viscosity}, the averages for $\alpha_{\rm vis}$ in the inner disks as reported in Tab.~\ref{tab:torusenergytab} are roughly constant among the different accretion disks explored here.

The extraction of effective $\alpha$-viscosities allows us to compute approximate viscous evolution timescales for our disks,
\begin{eqnarray}
	t_{\rm vis} &=& \frac{1}{\alpha}\left(\frac{R_0^{3}}{GM_{\rm BH}}\right)^{1/2}\left(\frac{z_H}{\varpi}\right)^{-2} \label{eq:tvisc} \\
	\approx&& \mskip-15mu 650\,{\rm ms} \left( \frac{\alpha}{0.01}\right)^{-1}\mskip-5mu\left(\frac{R_0}{30\,{\rm km}}\right)^{\frac{3}{2}} \mskip-5mu\left(\frac{M_{\rm BH}}{3M_\odot}\right)^{-\frac{1}{2}} \mskip-5mu\left(\frac{z_H/\varpi}{0.2}\right)^{-2}\mskip-5mu, \nonumber
\end{eqnarray}
where we have normalized to typical values extracted from our simulations in the second line. This estimate is remarkably similar to the duration of most short gamma-ray bursts, the bulk of which have typical durations of a few hundred milliseconds (as measured by their $T_{90}$, the time interval in which 5\% to 95\% of the total fluence is received by the detector; e.g., \citealt{berger_short-duration_2014,kumar_physics_2015}). The approximate viscous timescales for our simulation runs are listed in Tab.~\ref{tab:torusenergytab}; they span the range of short gamma-ray burst $T_{90}$ durations up to $T_{90}\lesssim 2\,{\rm s}$.

\subsubsection{Accretion}
\label{sec:accretion}

\begin{figure}[t]
  \includegraphics[width=\columnwidth]{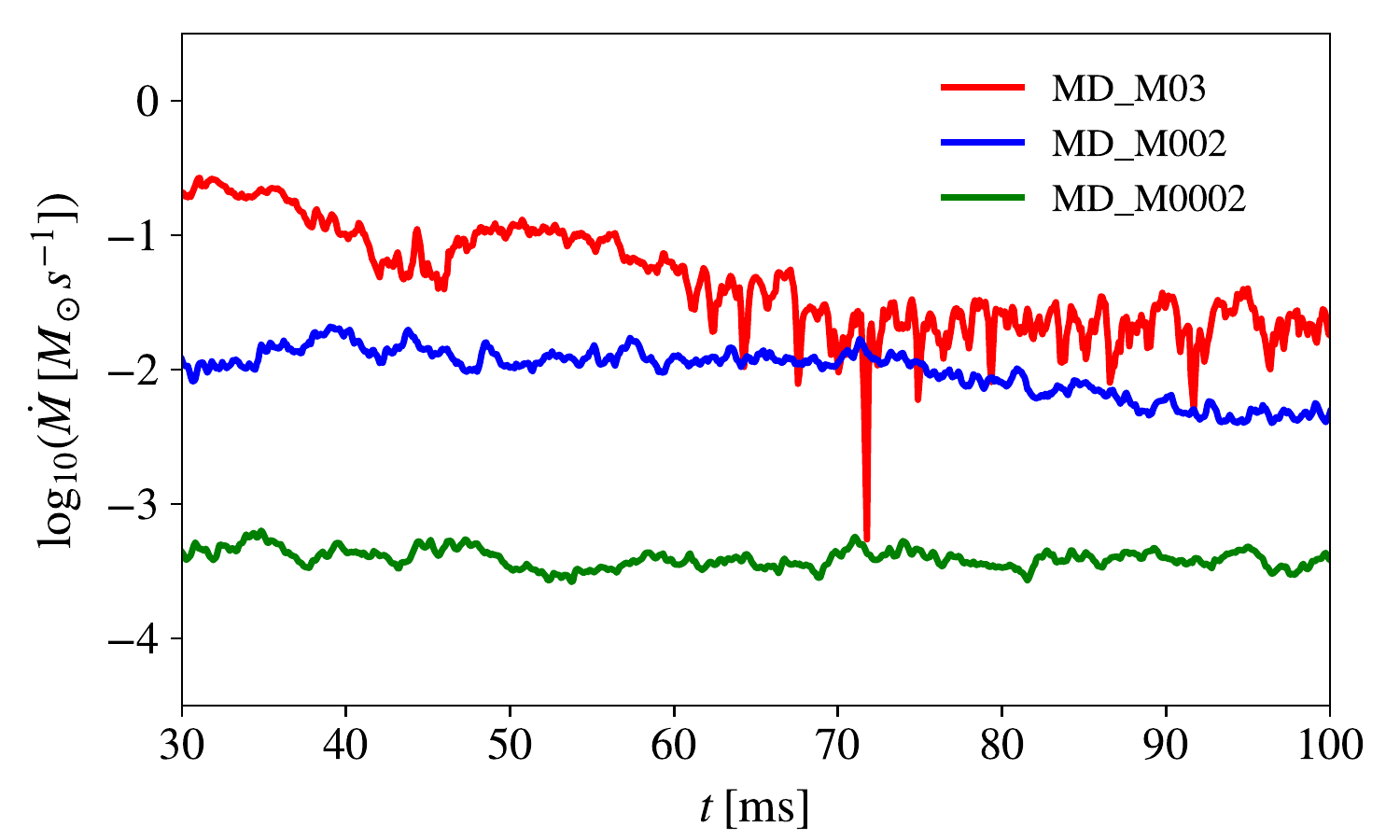}
  \includegraphics[width=\columnwidth]{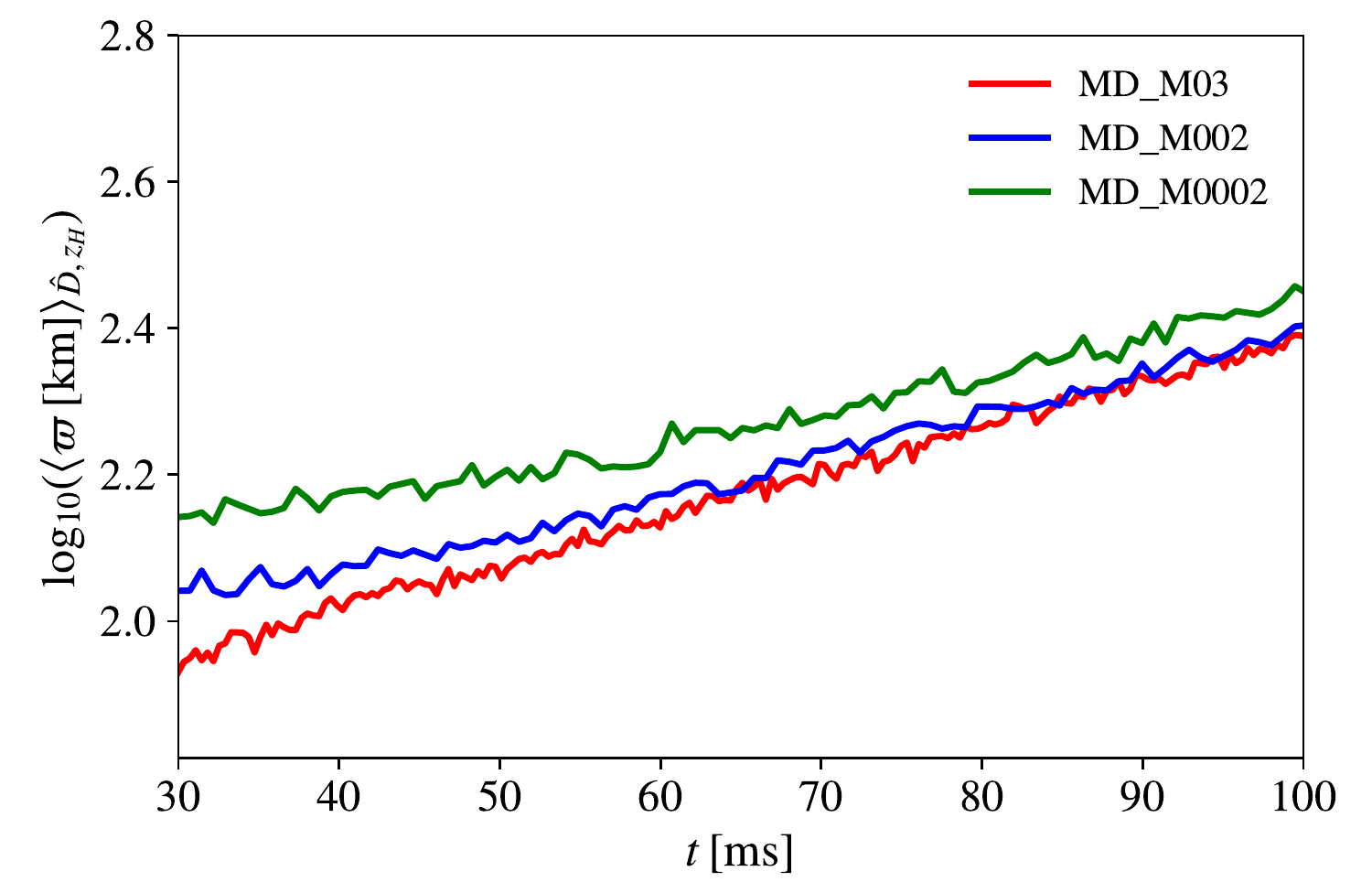}
 \caption{Top: Black hole accretion rates as a function of time for the three simulation runs in this work. Bottom: Density-averaged cylindrical radius of matter as a function of time, indicating viscous spreading over time. \label{fig:mdot-t}}
\end{figure}

\begin{figure}[t]
  \includegraphics[width=\columnwidth]{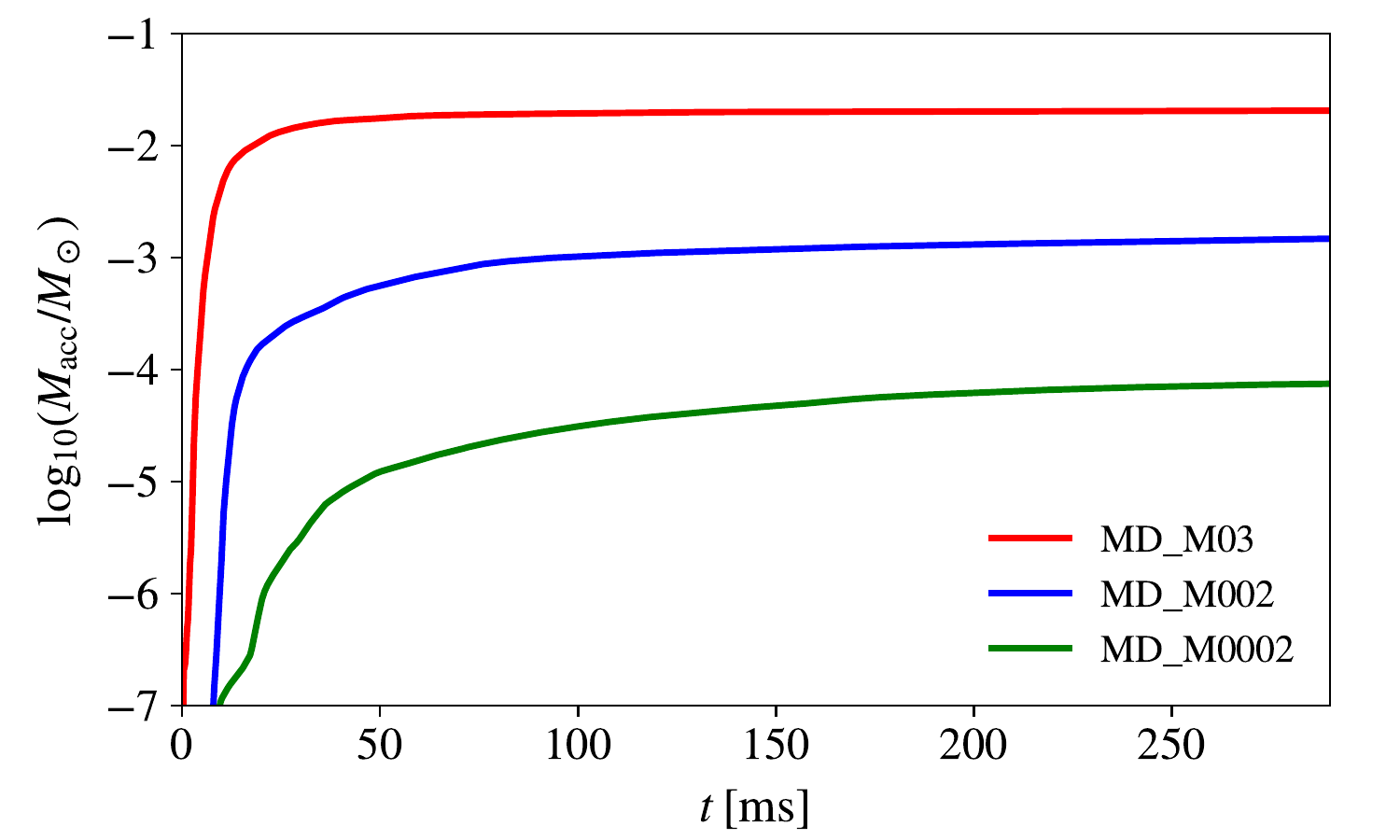}
 \caption{Total accreted mass onto the black hole as a function of time for the three simulation runs in this work. As a result of viscous spreading, the total cumulative accreted mass essentially converges over the duration of the simulation runs, and all remaining material will be eventually unbound from the system. \label{fig:macc-t}}
\end{figure}

Magnetic turbulence in the disks generated by the MRI drives accretion onto the black hole at the center, and outward transport of angular momentum in the disks. Figure \ref{fig:mdot-t} shows the evolution of accretion rates $\dot M$ with time for all three simulations. The accretion rates for models \texttt{MD\_M03}, \texttt{MD\_M002}, and \texttt{MD\_M0002}, time-averaged over a 10\,ms window around $t=30$\,ms are $2.6_{-0.7}^{+1.4}\times 10^{-1}~M_\odot s^{-1}$, 
$1.14_{-0.4}^{+0.4}\times 10^{-2}~M_\odot s^{-1}$, and $3.79_{-1.7}^{+2.5}\times 10^{-4}~M_\odot s^{-1}$ respectively; the accretion rate changes by a similar order of magnitude as the disk masses among the three simulation runs. This is expected from one-dimensional disk models, for which we have (see Eq.~\eqref{eq:Mdot_1D_2})
\begin{equation}
    \dot{M} \propto \alpha_{\rm vis} \bar{\rho}_{\rm d} M_{\rm BH}^2 \left(\frac{z_H}{\varpi}\right)^3, \label{eq:Mdot}
\end{equation}
where $\bar{\rho}_{\rm d}$ is the disk midplane density. We compute a representative value for $\bar{\rho}_{\rm d}$ in the inner accretion disk by calculating a three-dimensional spatial average of the disk following Eq.~\eqref{eq:rest_mass_dens_scaleht}, time-averaged over a 10\,ms window around $t=30$\,ms (see Tab.~\ref{tab:torusenergytab}). The average, minimum, and maximum values for the scale height $z_H/\varpi$ are extracted in an analogous way as for $\alpha_{\rm vis}$, restricting to the inner disk region between 30-175\,km in a 10~ms window centered at $t = 30$~ms. We find $0.23_{-0.1}^{+0.1}$ for \texttt{MD\_M03}, $0.23_{-0.1}^{+0.2}$ for \texttt{MD\_M002}, and $0.21_{-0.1}^{+0.2}$ for \texttt{MD\_M0002}. 
We note that the quantities, self-consistently set by MHD turbulence in our simulations, roughly satisfy relation \eqref{eq:Mdot}, within the uncertainties in extracting these numbers from the 3D turbulent evolution of the disks.

As a result of accretion onto the black hole, the outer parts of the disk are forced to viscously spread. The bottom panel of Fig.~\ref{fig:mdot-t} shows the early evolution of the density-averaged cylindrical radius of matter $\langle \varpi \rangle_{{\rm azim}, z_H}$ in the simulations, integrated up to the local density scale height. This parameter can serve as a rough indicator of radial viscous spreading; such spreading is indeed evident from Fig.~\ref{fig:mdot-t}. The smaller the viscous timescale (cf.~Tab.~\ref{tab:torusenergytab}) the faster the viscous spreading proceeds initially.

Figure \ref{fig:macc-t} shows the evolution of the accreted mass by the black hole as measured by the mass flux through a spherical coordinate detector surface placed at a radius of 12~km over the duration of the simulations. The total accreted mass by the black hole is $\approx$ 46\%, $\approx$ 60\%, and $\approx$ 31\% of the disk mass at 20~ms for \texttt{MD\_M03}, \texttt{MD\_M002}, and \texttt{MD\_M0002} respectively.  
Most of the accretion completes within the first $\sim$ 50-100~ms, after which the accretion rate starts to drop rapidly, with a comparatively negligible amount of mass projected to be accreted onto the black hole past the end of the simulation. We ascribe this effect to viscous spreading of the disks, which forces the disk material remaining at the end of the simulations to be unbound from the system. Properties of ejecta are discussed in Sec.~\ref{sec:ejecta}.

\subsubsection{Neutrino emission}
\label{sec:neutrino_emission}

\begin{figure*}[t]
  \includegraphics[width=\textwidth]{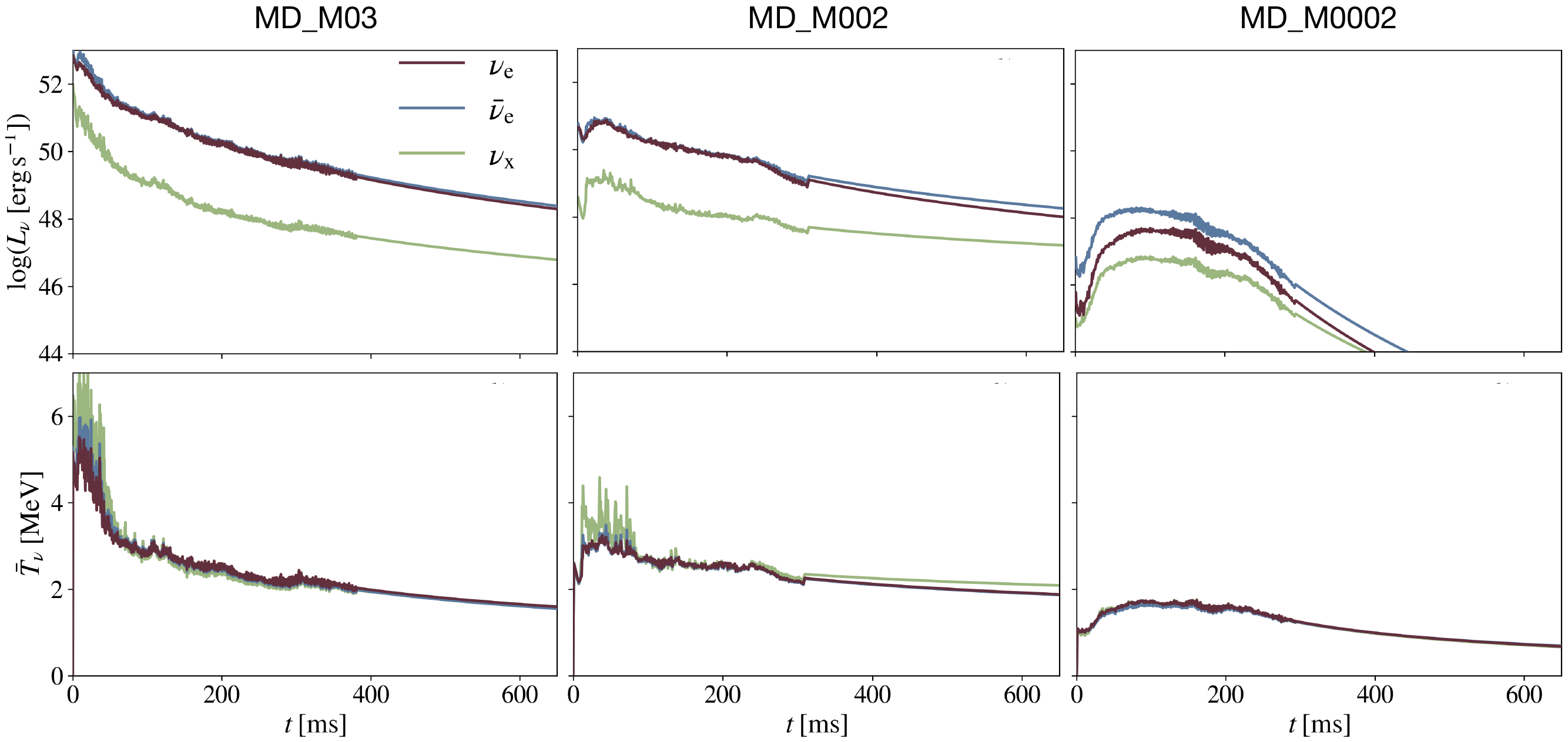}
 \caption{Characteristics of neutrino emission for the disk models explored in this work. Top: total neutrino luminosity. Bottom: mean neutrino emission temperature. The simulations end at $t \sim\!290$~ms after which the quantities are extrapolated by power law fits to late times.\label{fig:Lum_temp_nu}}
\end{figure*}

\begin{figure}[t]
  \includegraphics[width=\columnwidth]{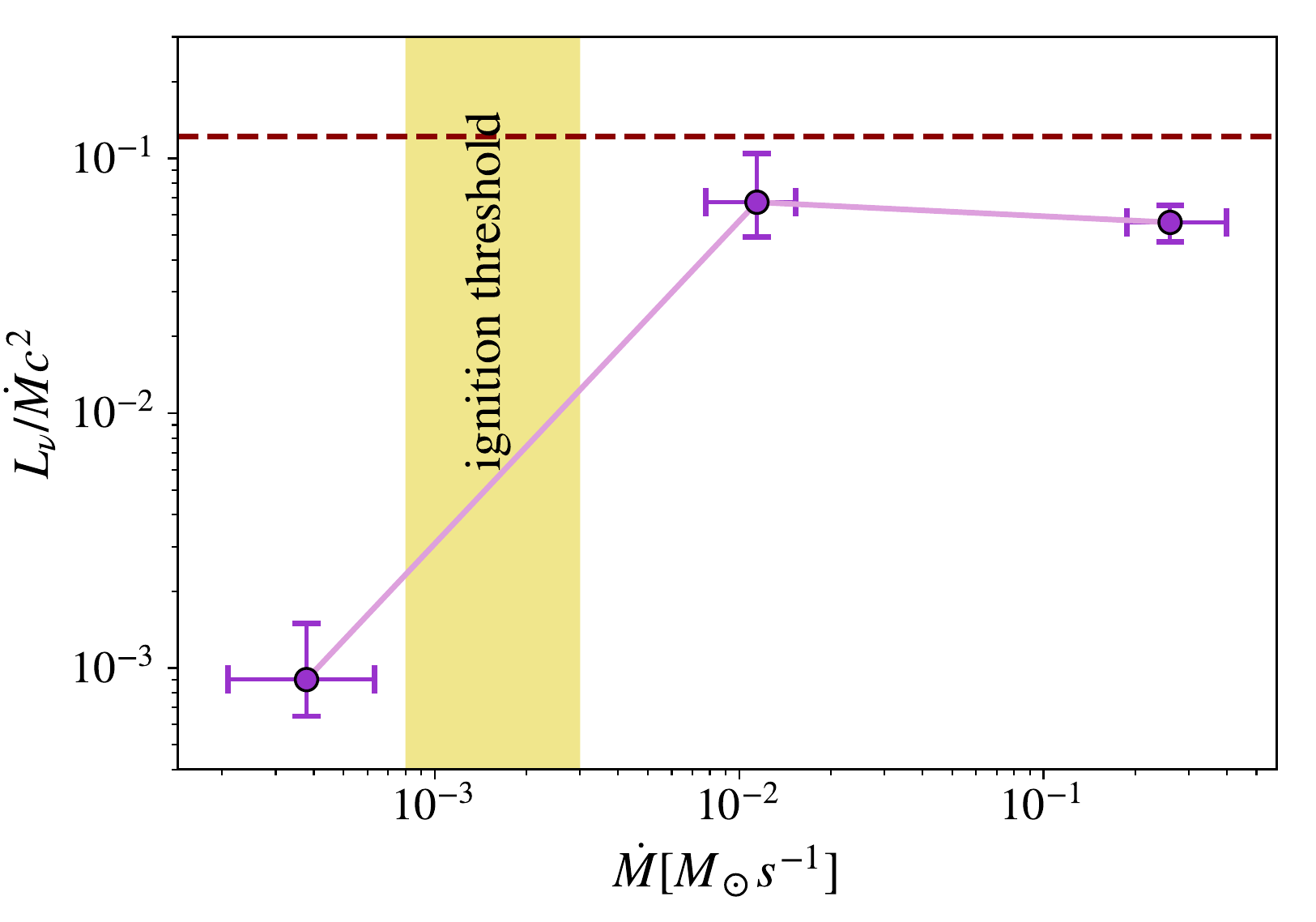}
 \caption{Radiative efficiency as a function of accretion rate for accretion onto a black hole with mass $3~M_\odot$ and dimensionless spin 0.8, as derived from the simulations in this work. The efficiency rises with increasing $\dot M$, reaches a maximum just above the ignition threshold $\dot{M}_{\rm ign}\sim 1\times 10^{-3}\,M_\odot\,{\rm s}^{-1}$ (cf.~Eq.~\eqref{eq:Mdot_ign_merger}), then decreases with increasing $\dot M$. The black dashed line shows the maximum radiative efficiency possible in such a system (see the text for details).\label{fig:lum-mdot}}
\end{figure}

Properties of neutrino radiation are reported in Figure \ref{fig:Lum_temp_nu} and Tab.~\ref{tab:torusenergytab}. As in \citet{siegel_three-dimensional_2018}, we define the total neutrino luminosity $L_\nu$ of a given species and the corresponding mean neutrino emission temperature $\bar T_\nu$ as
\begin{equation}
    L_{\nu_i} = \int \alpha W Q_{\nu_i}^{\rm eff} \alpha\sqrt{\gamma}d^{3}x,
\end{equation}
and
\begin{equation}
    \bar T_{\nu_i} = \frac{\int TQ_{\nu_i}^{\rm eff}W\alpha\sqrt{\gamma}d^{3}x}{\int Q_{\nu_i}^{\rm eff}W\alpha\sqrt{\gamma}d^{3}x},
\end{equation}
respectively. Here, $Q_{\nu_i}^{\rm eff}$ is the effective neutrino emissivity, $W$ is the Lorentz factor, and $\alpha$ is the lapse function. Neutrino species are labelled by $\nu_i \in {\nu_{e}, \bar \nu_{e}, \nu_{x}}$, where $\nu_{x}$ represents all heavier neutrino species collectively. We fit power laws to the late time simulation data to extrapolate these quantities beyond the time range modeled in the simulations. 

Figure \ref{fig:Lum_temp_nu} shows that the electron and anti-electron neutrino luminosities are at least an order of magnitude larger than the heavier neutrino luminosities. Emission channels for the heavier species are relatively suppressed at the comparatively low densities and temperatures of such accretion disks. For accretion disks above or close to the ignition threshold (see below and Sec.~\ref{sec:intro_ignition_threshold}), luminosities reach their maximum initially when the disks are still compact and in their high-density and high-temperature regime, before starting to viscously spread (cf.~\texttt{MD\_M03} and \texttt{MD\_M002}). For disks below the ignition threshold, neutrino luminosities may peak at a later time, once energy released through viscous accretion further heats up the gas and thus increases the neutrino emission (cf.~\texttt{MD\_M0002}, right panels of Fig.~\ref{fig:Lum_temp_nu}). Table \ref{tab:torusenergytab} reports the total $L_\nu$-value for $\nu_e$ and $\bar{\nu}_e$ for each simulation run, extracted as an average over a $10$\,ms time-window around the peak luminosities of the respective runs. These range between $\approx\!1\times 10^{52}$\,erg\,s$^{-1}$ (\texttt{MD\_M03}) and $\approx\!1\times 10^{48}$\,erg\,s$^{-1}$ (\texttt{MD\_M0002}).  
At later simulation times, as the accretion disks spread radially and become less compact, luminosities start to quickly fade over the timescales of the simulations. This indicates that neutrino self-irradiation of outflows in the context of r-process nucleosynthesis (see Sec.~\ref{sec:nucleosynthesis}) is likely only important initially, and much less so for the less luminous disks \texttt{MD\_M002} and \texttt{MD\_M0002}.

The qualitatively different behavior of our disk models in terms of weak interactions is captured by the differences in radiative efficiency $L_\nu/\dot M c^2$ among the simulations. Figure \ref{fig:lum-mdot} shows the variation of the radiative efficiency of the disks as a function of their accretion rate $\dot M$. The ratio $L_\nu/\dot M c^2$ represents the amount of accreted rest-mass energy that is turned into radiation per unit time. In order to assess radiative efficiency, for each simulation, we extract $L_\nu$ and $\dot M$ as mean values over the time range $t = 25-35$~ms. Fig.~\ref{fig:lum-mdot} shows the resulting efficiencies of $5.61^{+0.9}_{-0.9}\times 10^{-2}$, $6.73^{+3.7}_{-1.8}\times 10^{-2}$, and $9.01^{+6.0}_{-2.5}\times 10^{-4}$ for \texttt{MD\_M03}, \texttt{MD\_M002}, and \texttt{MD\_M0002}, respectively, compared to the maximum possible radiative efficiency. The latter is a fundamental limit on the amount of energy that can be extracted from a black hole accretion flow, determined by the available binding energy \citep{thorne_disk-accretion_1974}, 
\begin{equation}
    [L_\nu/\dot M c^2]_{\rm max} = 1 - E_{\rm ms},
\end{equation}
where
\begin{equation}
    E_{\rm ms} = \frac{1 - 2M_{\rm BH}}{3r_{\rm ms}}
\end{equation}
is the specific energy at the marginally stable circular orbit of a Kerr black hole \citep{bardeen_rotating_1972},
\begin{equation}
    r_{\rm ms} = M_{\rm BH}\{3 + Z_2 \mp [(3 - Z_1)(3 + Z_1 + 2Z_2)]^{1/2}\},
\end{equation}
with
\begin{eqnarray}
    Z_1 &=& 1 + (1 - \chi_{\rm BH}^2/M_{\rm BH}^2)^{1/3}[(1 + \chi_{\rm BH}/M_{\rm BH})^{1/3} \nonumber\\
    &&+ (1 - \chi_{\rm BH}/M_{\rm BH})^{1/3}], \\
    Z_2 &=& (3\chi_{\rm BH}^2/M_{\rm BH}^2 + Z_1^2)^{1/2}.
\end{eqnarray}

The efficiency in realistic scenarios, as also seen from the simulation data in Fig.~\ref{fig:lum-mdot}, is smaller than the maximum theoretical value, as a fraction of the binding energy is stored in the disk and radially advected into the black hole in the form of heat. The amount of heat radiated away thus depends on the efficiency of radiative processes with respect to radial advection of energy (cf.~Sec.~\ref{sec:ignition_threshold}). At low $\dot M$ values, the low midplane densities and temperatures in the inner disk suppress neutrino emission, resulting in a low radiative efficiency. As $\dot M$ increases, higher midplane densities and temperatures enhance neutrino emission, causing the radiative efficiency to rise and to reach a maximum just above the ignition threshold (Secs.~\ref{sec:intro_ignition_threshold} and \ref{sec:ignition_threshold}). As $\dot M$ increases further, neutrino cooling initially becomes more effective; eventually, however, the accretion timescale becomes shorter than the neutrino cooling timescale due to an increase in optical depth: the very high $\dot M$ values lead to high midplane densities (cf.~Eq.~\eqref{eq:Mdot}) and thus to an increase in the optical depth close to the midplane, eventually trapping neutrinos and reducing the cooling volume. One therefore expects enhanced radial advection of energy and a decrease in radiative efficiency with increasing accretion rate. 

This behavior is evident from Fig~\ref{fig:lum-mdot}, which shows a stark rise in radiative efficiency between the runs \texttt{MD\_M0002} and \texttt{MD\_M002} around an ignition threshold of $\dot{M}_{\rm ign}\sim 1\times 10^{-3}\,M_\odot\,{\rm s}^{-1}$ as predicted by Eq.~\eqref{eq:Mdot_ign_merger}. For even more massive accretion disks, a strong decline in the radiative efficiency would be expected as argued above (see also \citealt{chen_neutrino-cooled_2007}), but such disks are beyond the scope of the present study. We note that our results are qualitatively consistent with previous studies from one-dimensional disk models \citep{chen_neutrino-cooled_2007}.

\subsubsection{Ejecta}
\label{sec:ejecta}

\begin{figure*}[t]
  \includegraphics[width=\textwidth]{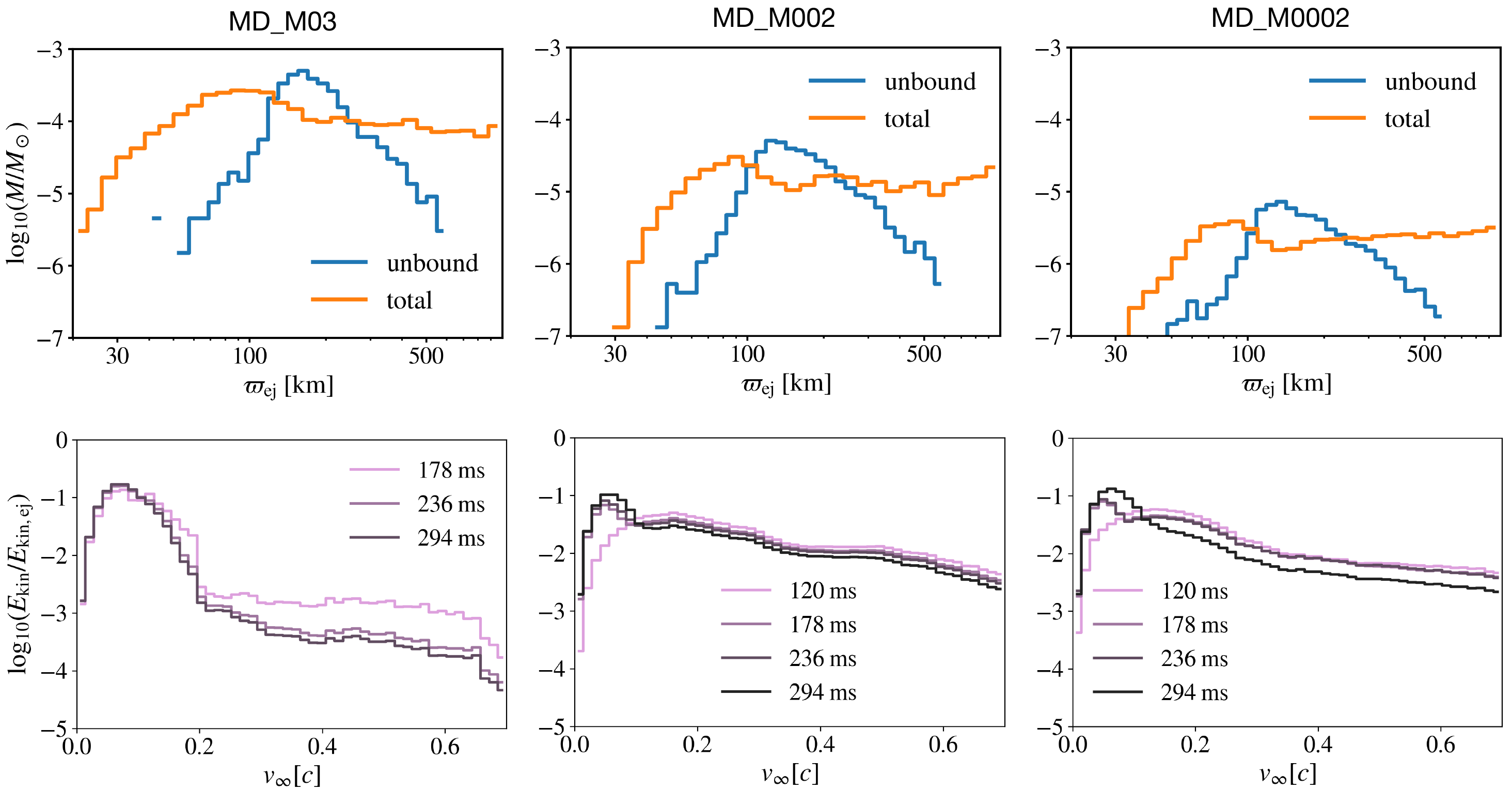}
 \caption{Outflow properties of the three accretion disks simulated here. Top: Mass distribution of unbound and total outflows as recorded by tracer particles in terms of their cylindrical radius $\varpi_{\rm ej}$ at which they get ejected from the disk corona. Bottom: Cumulative distribution of the kinetic energy of unbound outflows (normalized by the corresponding total kinetic energy) as a function of asymptotic escape velocity $v_\infty$ as estimated by a spherical detector surface placed at 1000\,km from the black hole and integrated up to different times.  \label{fig:outflows}}
\end{figure*}

Outflows from the disks originate in high specific entropy (`hot') disk coronae as a result of an imbalance between heating and cooling at high latitudes. Viscous heating as well as dissipation of magnetic energy far from the midplane is not offset by neutrino cooling, which becomes energetically subdominant at the comparatively low densities off the midplane. The precise amount of material unbound from the disk depends on the details of the self-consistent heating-cooling imbalance generated by MHD turbulence. We refer to \citet{siegel_three-dimensional_2018} for a detailed discussion on the emergence of these outflows in the presence of MHD turbulence.

We track properties of these outflows using $10^4$ tracer particles of equal mass, placed throughout the disk initially, with probability density proportional to the conserved rest mass density $\hat D = \sqrt \gamma \rho W$. The top panels of Fig.~\ref{fig:outflows} show the mass distribution of total and unbound disk outflow (ejecta) as recorded by the tracer particles in terms of the cylindrical radius at which the tracer particles are ejected from the disk corona. This ejection radius is defined as the cylindrical radius after which their radial coordinate position only increases with time. We define unbound material as having reached a coordinate radius of $10^3$\,km and additionally having positive specific energy at infinity, $-hu_0 > 0$, where $h$ denotes the specific enthalpy and $u_0$ is the covariant 0-component of the fluid four-velocity. For all simulations, we find that the total outflow material is almost evenly distributed over a broad  range of radii, whereas the majority of the ejecta mass originates from the inner accretion disk ($\varpi\lesssim100-250$~km), where most of the binding energy is released through viscous heating. The total amount of material ejected by the disks is estimated to be $3.179\times 10^{-3}~M_\odot (\approx 17\%$ of the initial disk mass) 
for \texttt{MD\_M03}, $4.03\times 10^{-4}~M_\odot (\approx 22\%$ of the initial disk mass) 
for \texttt{MD\_M002}, and $6.046\times 10^{-5}~M_\odot (\approx 30\%$ of the initial disk mass) 
for \texttt{MD\_M0002} by the end of the corresponding simulations.

The disks still generate steady winds at the end of our simulations, which indicates that the remaining disks would continue evaporating themselves if the simulations were evolved over longer timescales. Given that the total cumulative accreted masses have already converged over the duration of the simulation runs (cf.~Sec.~\ref{sec:accretion}), we predict that the remaining disk will generate outflow material corresponding to $\approx\! 37\%$ of the initial disk mass for \texttt{MD\_M03}, $\approx\! 18\%$ of the initial disk mass for \texttt{MD\_M002}, and $\approx\! 38\%$ of the initial disk mass for \texttt{MD\_M0002}. 
This material will be ejected over longer timescales by a combination of the MHD-driven outflows as discussed here and viscous spreading of the disk over several viscous timescales (see \citealt{fernandez_long-term_2019} for a discussion of late-time viscous outflows). Ejecta percentages are uncertain by at least $\pm 10\%$ due to the ambiguities and approximations in defining and measuring ejected mass.

While we find disks above the ignition threshold (cf.~Secs.~\ref{sec:intro_ignition_threshold} and \ref{sec:neutrino_emission}) eject roughly 30--50\% of their initial disk mass after relaxation, disks below the ignition threshold even eject more $\gtrsim 60\%$. We ascribe such increased evaporation to the absence of a significant cooling mechanism for the accretion flow not just off the midplane, but also in the midplane and thus throughout the disk. Strongly enhanced heating is evident both in the midplane (cf.~the large midplane entropy for \texttt{MD\_M0002} compared to \texttt{MD\_M03} and \texttt{MD\_M002} in Fig.~\ref{fig:panels_R_eff_entropy}) as well as in the `corona' (cf.~the increase in specific entropy at high latitudes between \texttt{MD\_M0002} and \texttt{MD\_M002} in Fig.~\ref{fig:entropy_space-time}).

The bottom panels of Figure \ref{fig:outflows} show the distribution of kinetic energy of the disk ejecta in terms of the asymptotic escape speed $v_\infty$ as measured by a spherical detector surface placed at $\approx\!1000$\,km from the black hole. The asymptotic velocities $v_\infty$ are computed from the asymptotic Lorentz factor $W_\infty = -hu_0$. In all three simulations, we find the emergence of a high-velocity tail, with the bulk ejecta residing around $v_\infty\approx 0.1c$. We find kinetic-energy weighted mean escape velocities of $\bar{v}_{\infty}\approx 0.14c,\,0.21c,\,0.17c$ for \texttt{MD\_M03}, \texttt{MD\_M002}, and \texttt{MD\_M0002}, respectively. This velocity scale is naturally explained as a combination of moderate outflow velocities from the hot corona of typically $(0.03-0.1)c$ together with the energy released by recombination of individual nucleons into $\alpha$-particles ($\approx\!7$\,MeV per baryon per $\alpha$-particle formed). Our velocity distributions are qualitatively similar to \citet{fernandez_long-term_2019} and \cite{Christie:2019lim}, albeit their GRMHD run---similar to our \texttt{MD\_M03} case---shows more outflow mass at higher velocities; this has been commented on in \citet{fernandez_long-term_2019} and can be attributed in part to their initial magnetic field configuration, which is optimized for fast magnetically-dominated outflows. The high-velocity tail for the low-$\dot{M}$ disks \texttt{MD\_M002} and \texttt{MD\_M0002} are more pronounced, which may be ascribed to more violent viscous heating as neutrino cooling becomes less important at low accretion rates (cf.~increase in specific entropy in Figs.~\ref{fig:panels_R_eff_entropy} and \ref{fig:entropy_space-time}).

\subsection{Weak interactions and disk composition}
\label{sec:disk_composition}

\begin{figure*}[t]
  \includegraphics[width=\textwidth]{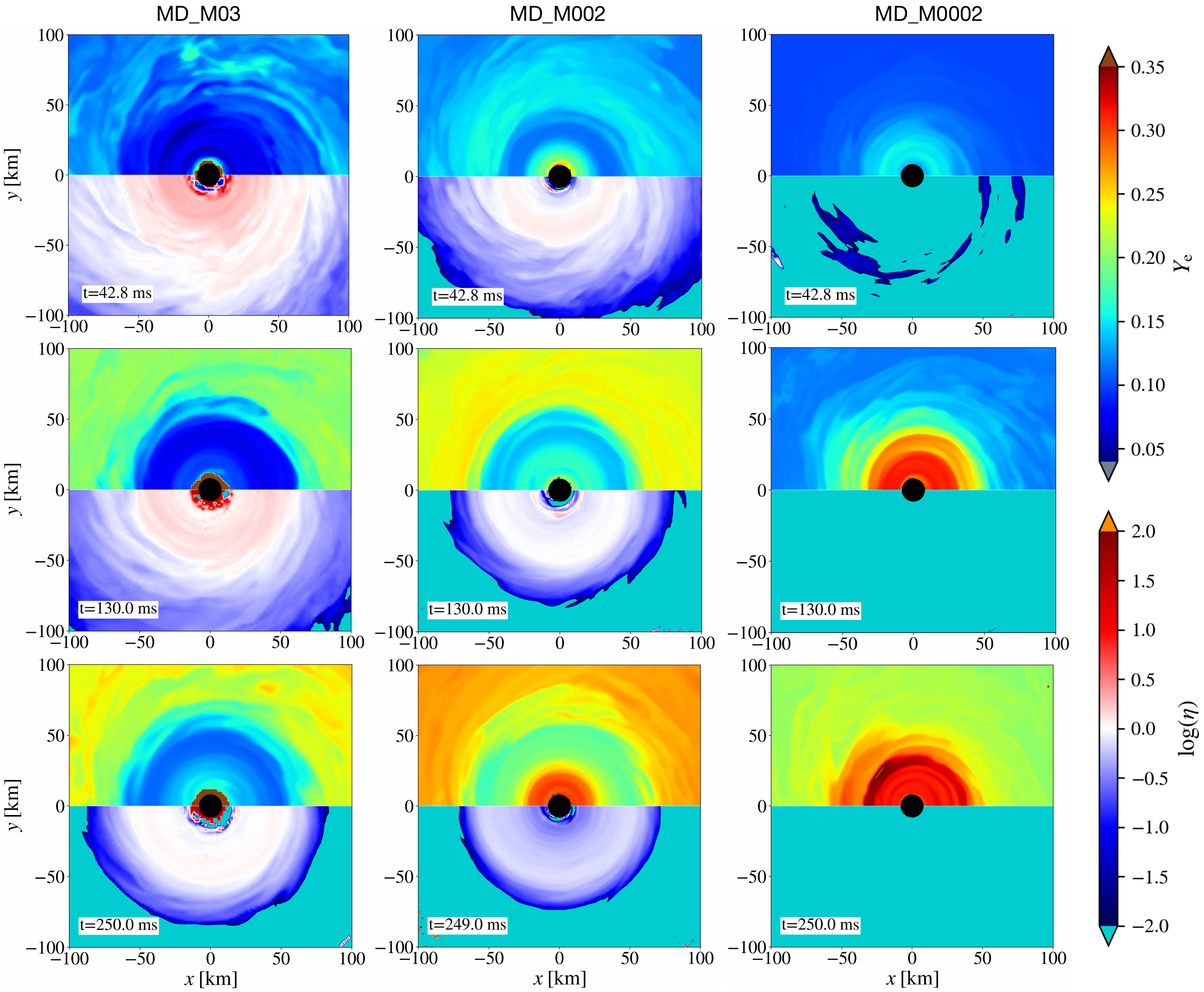}
 \caption{Snapshots of the electron fraction $Y_e$ and electron degeneracy $\eta=\mu_e/k_{\rm B}T$ over time for the three simulation runs performed here. Plotted are slices of these quantities in the disk midplane ($x$-$y$ plane) at times $t = 42.8$~ms, $t = 130$~ms, and $t = 250$~ms. Above the ignition threshold for weak interactions (Eq.~\eqref{eq:Mdot_ign_merger}), the high accretion rates of runs \texttt{MD\_M03} and \texttt{MD\_M002} cause the inner disk to remain strongly neutron-rich over time, whereas below the ignition threshold the low accretion rate of run \texttt{MD\_M0002} leads to accelerated protonization in the inner part. \label{fig:panels_Ye_eta}}
\vspace{5mm}
\end{figure*}

\begin{figure*}[t]
\centering
  \includegraphics[width=\textwidth]{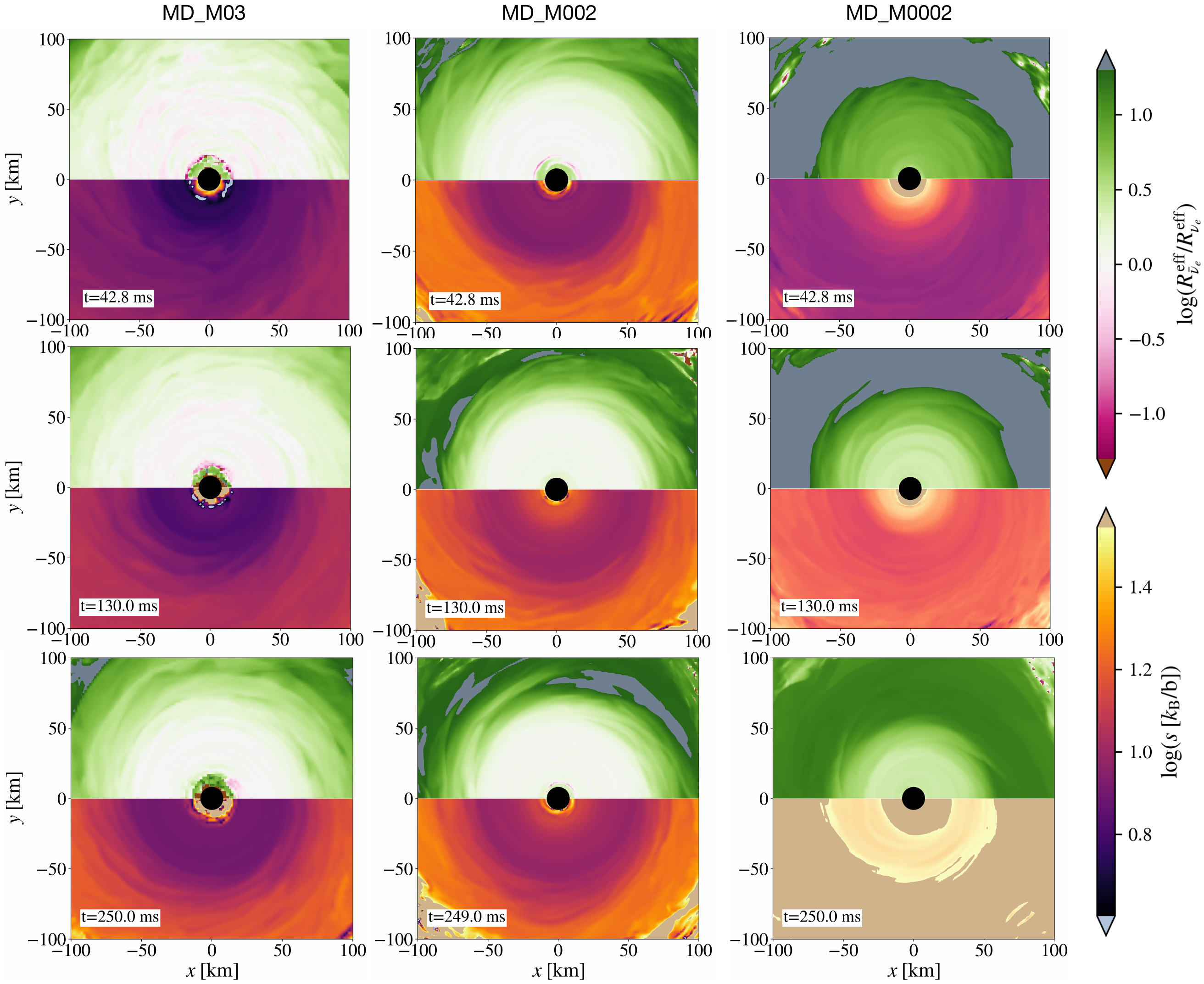}
 \caption{Evolution of the ratio of anti-electron to electron neutrino number emission rates $R_{\bar \nu_{\rm e}}^{\rm eff}/R_{\nu_{\rm e}}^{\rm eff}$ and specific entropy $s$ for the three simulation runs performed here. Plotted are snapshots of these quantities in the disk midplane ($x$-$y$ plane) at times $t = 42.8$~ms, $t = 130$~ms, and $t = 250$~ms.\label{fig:panels_R_eff_entropy}}
\vspace{5mm}
\end{figure*}

\begin{figure}[t]
  \includegraphics[width=\columnwidth]{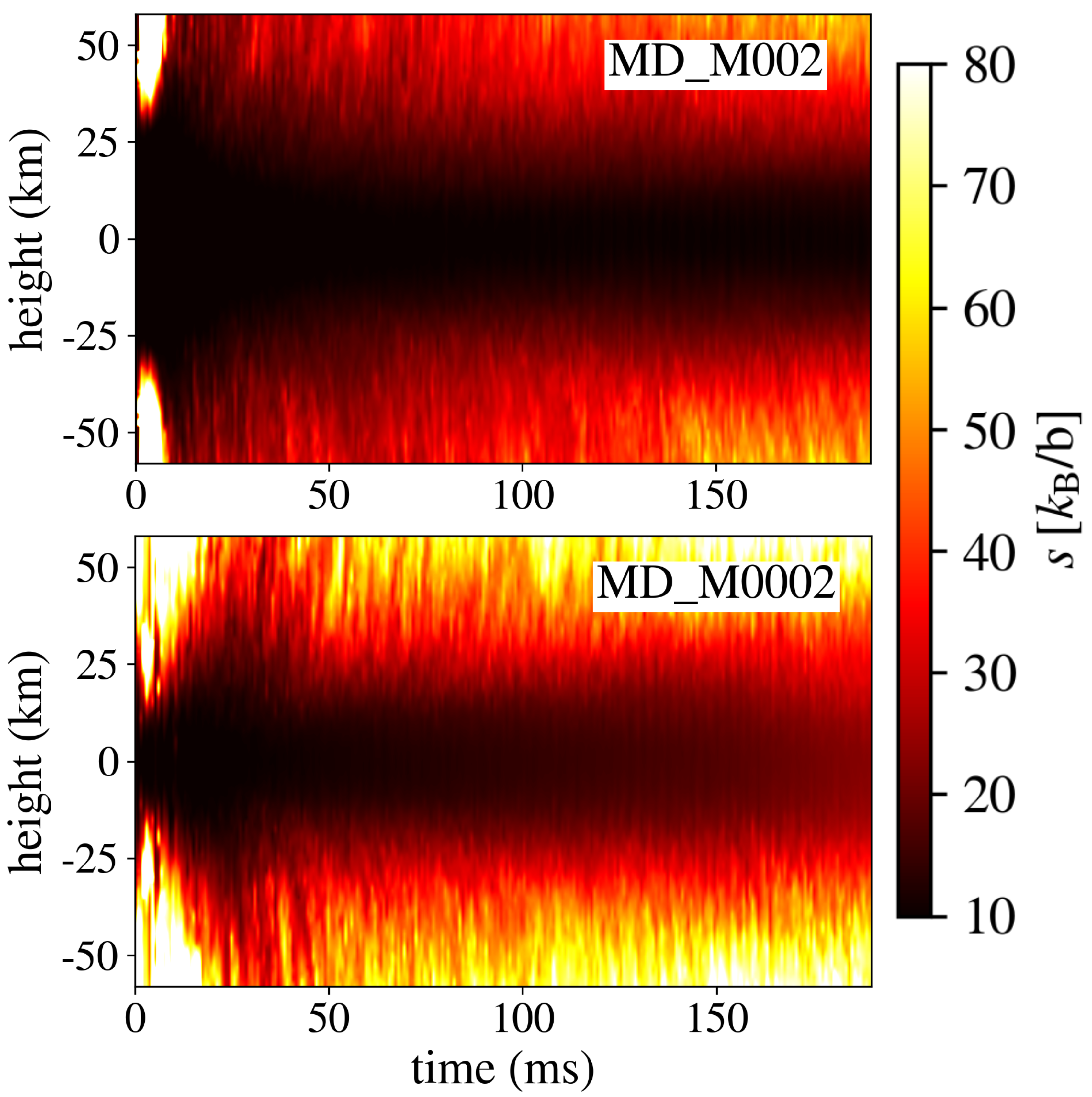}
 \caption{Space-time diagram of specific entropy in accretion disks above (top) and below (bottom) the ignition threshold. For each disk, variation of specific entropy with height (in the $z$ direction) measured from the disk midplane ($z = 0$) is shown as a function of time; at each epoch, specific entropy values are extracted from the $x-z$ plane of the disk with a radial averaging performed between 45\,km and 70\,km.\label{fig:entropy_space-time}}
\vspace{5mm}
\end{figure}

\begin{figure*}[t]
  \includegraphics[width=\textwidth]{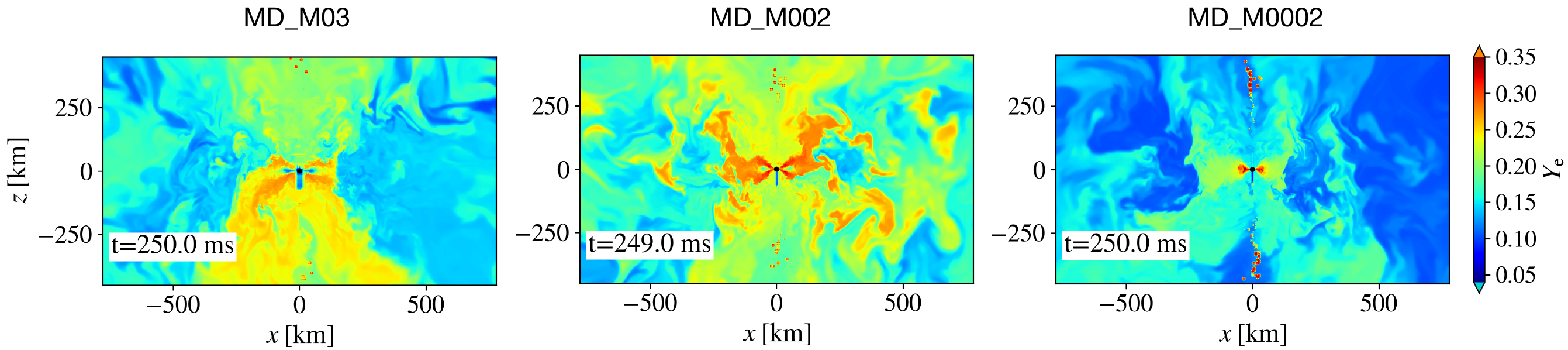}
 \caption{Snapshots of the electron fraction $Y_e$ at $t = 250$~ms in the meridional plane ($x$-$z$ plane) for the three simulation runs performed here, indicating lateral as well as radial composition gradients and significant mixing. \label{fig:panels_Ye_xz}}
\vspace{5mm}
\end{figure*}

Accretion disks below and above the ignition threshold for weak interactions give rise to qualitatively different evolution of their composition as we shall illustrate in this section. Discussing weak interactions in the disk in more detail also benefits the interpretation of the behavior of some global disk quantities (Sec.~\ref{sec:global_properties}) as well as of nucleosynthesis in disk outflows (Sec.~\ref{sec:nucleosynthesis}).

Figures \ref{fig:panels_Ye_eta} and \ref{fig:panels_R_eff_entropy} show various quantities pertaining to the evolution of disk composition, including snapshots of the electron fraction, electron degeneracy $\eta=\mu_e/k_{\rm B}T$, where $\mu_e$ is the electron chemical potential, the ratio of $\bar{\nu}_e$ to $\nu_e$ number emission rates ($R_{\bar \nu_{\rm e}}^{\rm eff}/R_{\nu_{\rm e}}^{\rm eff}$), and specific entropy $s$ over the course of the three simulation runs.

Neglecting absorption of neutrinos, roughly appropriate for the disk midplane in the regimes probed here, the weak interactions controlling disk composition are the charged-current $\beta$-processes,
\begin{eqnarray}
      e^- + p &\rightarrow& n + \nu_{\rm e}, \label{eq:proton_capture} \\
      e^+ + n &\rightarrow& p + \bar \nu_{\rm e}. \label{eq:neuton_capture}
\end{eqnarray}
Starting off from neutron-rich initial conditions $Y_e\approx 0.1$ (cf.~Sec.~\ref{sec:methods}), one expects the disks to protonize over time due to the dominant reaction Eq.~\eqref{eq:neuton_capture}. This is evident from $R_{\bar \nu_{\rm e}}^{\rm eff}/R_{\nu_{\rm e}}^{\rm eff}>0$ and the gradual increase of $Y_e$ over time in the outer parts of the accretion disks in Figs.~\ref{fig:panels_R_eff_entropy} and \ref{fig:panels_Ye_eta}. However, less massive disks with lower $\dot{M}$ such as \texttt{MD\_M002} and \texttt{MD\_M0002} take longer times compared to \texttt{MD\_M03} to raise their electron fraction from its initial value; this can be understood from an analysis of the disk protonization timescale.

From the equations of lepton number and baryon number conservation, we
compute
\begin{equation}
  R = \nabla_\mu(n_e u^\mu) = \nabla_\mu(n_{\rm b} u^\mu Y_e) = n_{\rm b} u^\mu
  \nabla_\mu Y_e,
\end{equation}
and thus
\begin{equation}
  u^\mu \nabla_\mu Y_e = R \frac{m_{\rm b}}{\rho}, \label{eq:Y_e_cons}
\end{equation}
where $u^\mu$ is the four-velocity of the fluid and $R$ denotes a source term to account for the change in lepton number due to weak interactions. Equation \eqref{eq:Y_e_cons} shows that along a fluid trajectory, i.e., in the comoving frame of the fluid, the electron fraction changes by a rate of $R m_{\rm b}/ \rho$. We can therefore define the characteristic timescale for the composition of matter to change by $t_{Y_e} \equiv Y_e\rho / (R m_\mathrm{b})$. Neglecting neutrino absorption, appropriate for the disk midplane, one has
\begin{equation}
  t_{Y_e} = \frac{Y_e}{R_{\bar{\nu_e}}^\mathrm{eff} -
    R_{\nu_e}^\mathrm{eff}}\frac{\rho}{m_{\rm b}}. \label{eq:t_Ye}
\end{equation}
Ignoring final state blocking in the neutrino phase space, the effective emission rates scale as \citep{tubbs_neutrino_1975,bruenn_stellar_1985} $R_{\bar{\nu}_e,\nu_e}^\mathrm{eff} \propto \rho T^5 \propto \dot{M}^{9/4}$, where we have used the approximate expression Eq.~\eqref{eq:T_1D} for the midplane temperature in the second step. Furthermore, since $\dot{M}\propto \rho$ (cf.~Eq.~\eqref{eq:Mdot_1D_2}), we deduce that, approximately,
\begin{equation}
  t_{Y_e} \propto \dot{M}^{-5/4},  \label{eq:t_Ye_scale_Mdot}
\end{equation}
which explains the decrease in protonization timescale with increasing accretion rate seen in the outer accretion disks in Figs.~\ref{fig:panels_Ye_eta} and \ref{fig:panels_R_eff_entropy}.

The inner accretion disk shows qualitatively different behavior depending on whether the accretion disk resides in a state above or below the ignition threshold (Eq.~\eqref{eq:Mdot_ign_merger}, Sec.~\ref{sec:neutrino_emission}). We find that above the ignition threshold, the disk midplane density $\rho\propto \dot{M}$ (cf.~Eq.~\eqref{eq:Mdot_1D_2}) is sufficiently large that electrons become degenerate, as shown in Fig.~\ref{fig:panels_Ye_eta}. This, in turn, suppresses positron creation via $\gamma\rightarrow e^+ + e^-$ and thus Eq.~\eqref{eq:neuton_capture} relative to \eqref{eq:proton_capture}, and hence leads to slight neutronization even in very neutron-rich initial conditions (see \texttt{MD\_M03} in Fig.~\ref{fig:panels_R_eff_entropy}). This neutronization mechanism is also why collapsar accretion disks starting with much higher initial $Y_e$ are able to generate neutron-rich outflows and synthesize r-process elements \citep{siegel_collapsars_2019}. However, the inner part of the accretion disks cannot become arbitrarily degenerate and neutron rich, thanks to a self-regulation mechanism discussed in \citet{siegel_three-dimensional_2018} and first pointed out by \citet{beloborodov_nuclear_2003} (see also \citet{chen_neutrino-cooled_2007}). As shown in Fig.~\ref{fig:panels_Ye_eta}, disk self-regulation leads to a heating-cooling balance that results in moderate electron degeneracy $\eta\sim 1$ and corresponding neutron richness of $Y_e\approx 0.1$. While the disk is in such a self-regulated, moderately degenerate phase, the inner part of the accretion disk feeds highly neutron-rich material into the outflows. At lower $\dot{M}$, close to the ignition threshold, degeneracy and self-regulation become somewhat weaker and less pronounced; however, \texttt{MD\_M002} still qualitatively shows the same behavior as \texttt{MD\_M03}. 

As a result of viscous spreading, the disk density decreases over time, which may suppress degeneracy eventually late in the disk's evolution. The onset of a decrease in degeneracy is noticeable from the snapshots in Fig.~\ref{fig:panels_Ye_eta}. However, by the time massive disks such as \texttt{MD\_M03} reach this break-down of degeneracy, most of the ejecta material has already been launched and the change in disk composition has little effect on the overall composition of ejecta.

Below the ignition threshold, we find an inverted scenario in terms of midplane composition in the inner parts of the accretion disk. The timescale for protonization in the innermost part of the accretion disk is significantly smaller than in the outer parts, resulting in high-$Y_e$ material surrounding the black hole already at $t\sim100$\,ms (cf.~\texttt{MD\_M0002} in Fig.~\ref{fig:panels_Ye_eta}). We attribute this to excess viscous heating in the absence of energetically significant neutrino cooling. The resulting high-entropy environment (cf.~Figs.~\ref{fig:panels_R_eff_entropy} and \ref{fig:entropy_space-time}) leads to a prolific generation of positrons via pair production, $\gamma\rightarrow e^+ + e^-$, and thus decreases the timescale for protonization via Eq.~\eqref{eq:neuton_capture}. This innermost region in \texttt{MD\_M0002} is too small, however, to feed material into the outflows sufficient enough to form a pronounced high-$Y_e$ tail of the ejecta (cf.~Sec.~\ref{sec:nucleosynthesis}).

Figure \ref{fig:panels_Ye_xz} shows a snapshot of $Y_e$ in the meridional plane on large spatial scales $\lesssim 700$\,km. In general, the proton fraction is enhanced in polar outflows, in agreement with enhanced pair production in the low-density polar funnels (Fig.~\ref{fig:optical_depth}). A radial $Y_e$ gradient is also present, which results from the gradual protonization of the outer disks over time according to Eq.~\eqref{eq:t_Ye_scale_Mdot}. As a result of a long `engine lifetime' on a timescale $0.1-1$\,s, continuous ejection of wind material leads to significant mixing of outflow material before it enters a homologous expansion phase. 
Three-dimensional radiation transport calculations will be required to compute detailed predictions for kilonova lightcurves and spectra.

\subsection{Nucleosynthesis}
\label{sec:nucleosynthesis}

\begin{figure}[t]
  \includegraphics[width=\columnwidth]{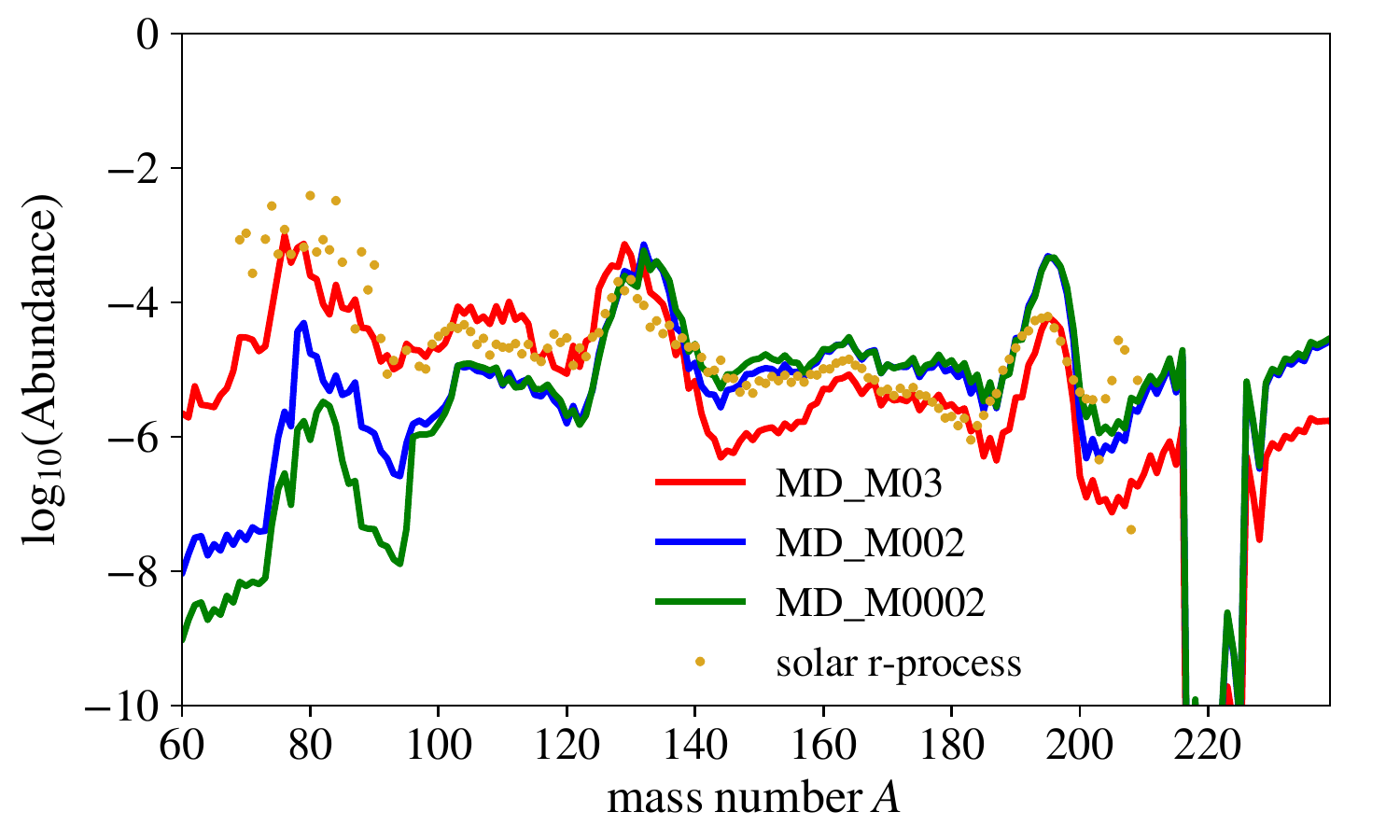}
 \caption{Final elemental abundances at $10^9$\,s for the accretion disk ejecta of our three simulation runs. Shown are mean abundances of all unbound tracer particles (excluding the disk relaxation phase). Also plotted for reference are the observed solar system abundances from \citet{arnould_$r$-process_2007}, scaled to match the \texttt{MD\_M03} abundance at $A = 195$. \label{fig:abundances}}
\end{figure}

\begin{figure}[t]
  \includegraphics[width=\columnwidth]{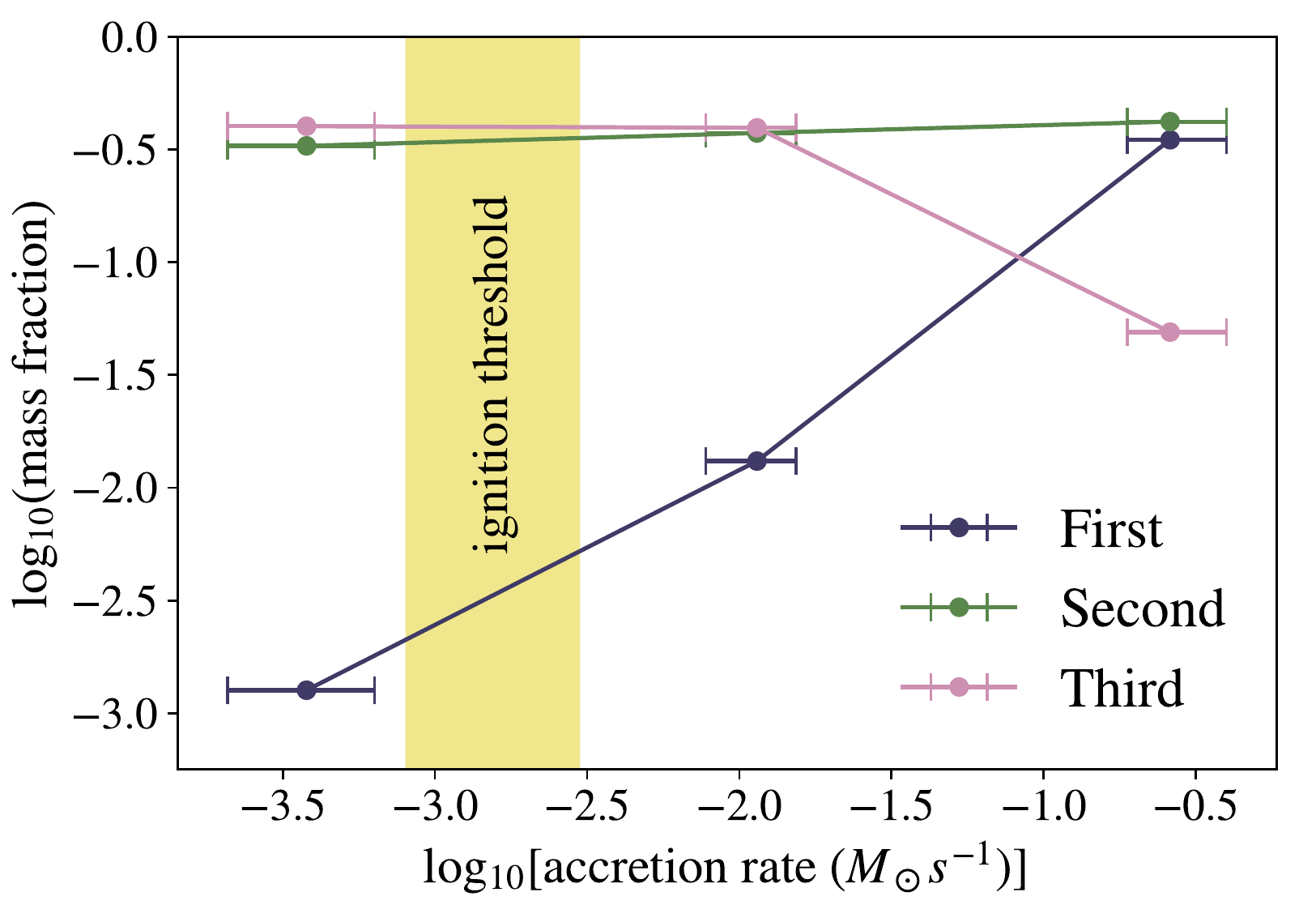}
 \caption{Mass fractions of the nucleosynthetic yields in
the first (summed over $A=70-90$), second (summed over $A=125-135$), and third (summed over $A=186-203$) r-process peaks for the disk ejecta of simulation runs \texttt{MD\_M03}, \texttt{MD\_M002}, and \texttt{MD\_M0002} presented here. A qualitative change in nucleosynthesis products across the ignition threshold for weak interactions (Eq.~\eqref{eq:Mdot_ign_merger}) is evident. \label{fig:mass_frac_peaks}}
\end{figure}

\begin{figure}[t]
  \includegraphics[width=\columnwidth]{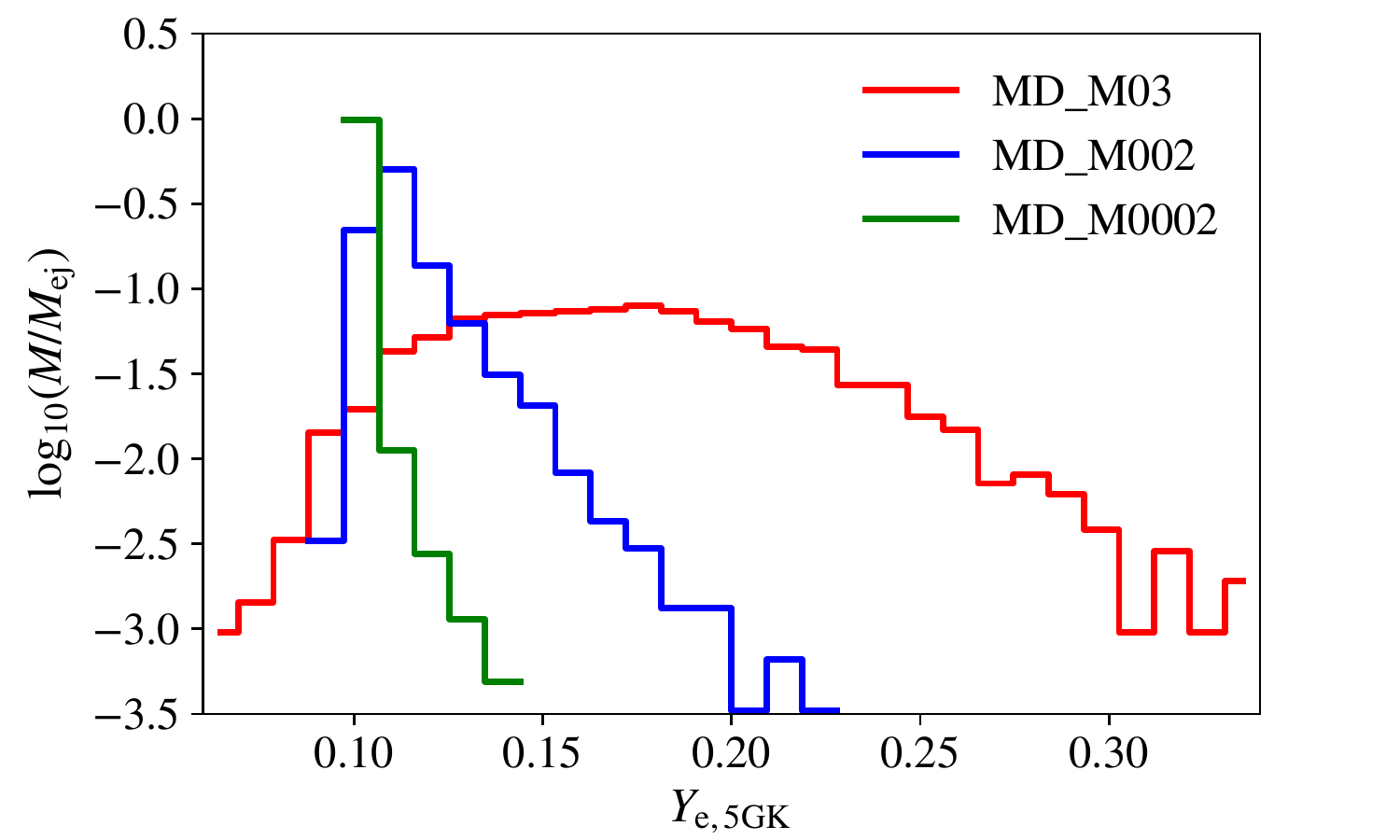}
 \caption{Mass distribution of unbound outflows from our simulated accretion disks according to the electron fraction $Y_{\rm e}$, extracted from tracer particles at 5\,GK. A significant high-$Y_e$ tail is established across the ignition threshold. \label{fig:mass-Ye-outflow}}
\end{figure}

The neutron-rich outflows generated by our post-merger accretion disks are sites for r-process nucleosynthesis. We record thermodynamic properties of the ejecta by tracer particles distributed throughout the disks initially (cf.~Sec.~\ref{sec:ejecta}), and use such tracer profiles as input to nuclear reaction network calculations. The initial conditions that determine the outcome of the r-process are mostly set by the electron fraction, specific entropy, and expansion timescale of the flow at $T\approx 5$\,GK \citep{lippuner_r-process_2015}, which is the characteristic temperature at which nuclear statistical equilibrium breaks down and neutron-capture reactions set in.

Neutrino absorption is taken into account approximately by a ring-like `light-bulb' scheme in post-processing \citep{fernandez_delayed_2013}, which irradiates the ejecta with neutrino luminosities as computed in Sec.~\ref{sec:neutrino_emission}. We refer to \citet{fernandez_delayed_2013} and \citet{siegel_three-dimensional_2018} for more details on this approach.

Starting at 10~GK, we perform full nuclear reaction network calculations on the tracer particles using the reaction network \texttt{SkyNet}~\citep{lippuner_skynet:_2017}, in order to track nuclear abundances as the outflows undergo r-process nucleosynthesis. Figure \ref{fig:abundances} shows the final total abundance yields from all tracers after $10^9$\,s for all three disk simulation runs. These abundances are the result of integrated ejecta that are unbound over the course of the simulation runs, most of which originate from the inner accretion disk region with radii between $\approx(100-250)$~km (cf.~Fig.~\ref{fig:outflows}). 

Final r-process abundance patterns qualitatively change across the ignition threshold for weak interactions in post-merger disks (see Fig.~\ref{fig:abundances}). At high accretion rates, such as run \texttt{MD\_M03}, we find abundances in good agreement with residual solar r-process abundances~\citep{arnould_$r$-process_2007} across the entire range of mass numbers from the first ($A\approx80$) to the third ($A\approx195$) r-process peak (see \citet{siegel_three-dimensional_2018} for a more detailed discussion on this case). While across all accretion rates we find good agreement with the solar abundance pattern between the second peak ($A\approx130$) and the third r-process peak, run \texttt{MD\_M002} (close to the ignition threshold) already shows somewhat suppressed first-to-second peak elements, and light r-process elements are strongly suppressed below the ignition threshold (run \texttt{MD\_M0002}).

The qualitative change in nucleosynthesis patterns across the ignition threshold is further illustrated by Fig.~\ref{fig:mass_frac_peaks}, which shows the mass fractions of nuclei grouped into first-, second-, and third-peak elements as a function of accretion rate. While 2nd-peak elements are synthesized in roughly constant amounts across the $\dot{M}$ range investigated here, abundances of light r-process nuclei rise strongly toward and around the ignition threshold, while 3rd-peak nuclei are somewhat overproduced in that regime.

This behavior is a result of a qualitative change in the distribution of $Y_e$ of the ejecta across the ignition threshold as evident from Fig.~\ref{fig:mass-Ye-outflow}. While at low $\dot{M}$, below the ignition threshold, the $Y_e$ distribution of the unbound outflows at 5\,GK is still centered around the initial value of $Y_e=0.1$ (run \texttt{MD\_M0002}), run \texttt{MD\_M002} close to the ignition threshold has already built up a significant tail toward higher electron fraction; finally, run \texttt{MD\_M03} shows a broad distribution of $Y_e$ at least up to $\lesssim 0.4$. The emerging high $Y_e$ tail with increasing disk masses across the ignition threshold was also reported by \citet{Metzger:2008jt}. This effect can be  ascribed in part to the strongly changing protonization timescale as a function of accretion rate across the ignition threshold (cf.~Eq.~\eqref{eq:t_Ye_scale_Mdot}; Sec.~\ref{sec:disk_composition}). There is a possibility of our tracer particles slightly under-resolving the high-$Y_e$ tail---a significantly larger number of tracers, not considered here due to computational cost and feasibility, may better resolve the effects of, e.g., an inner high-$Y_e$ part as in the \texttt{MD\_M0002} disk, but is unlikely to qualitatively change the conclusions reached here. Toward high $\dot{M}$, neutrino absorption by the ejecta contributes to flattening the $Y_e$ distribution as well. This effect is already noticeable for \texttt{MD\_M03} and is expected to become more important at even higher-$\dot{M}$ disks \citep{miller_full_2019-1}, which are beyond the scope of the present paper.

\section{Discussion \& Conclusions}
\label{sec:conclusion}

Our simulation results on post-merger accretion disks address a number of key questions regarding r-process nucleosynthesis and the electromagnetic counterpart observables of LIGO-Virgo's BNS and NSBH merger source populations. In view of future gravitational-wave detections, we explore the parameter space of BNS and NSBH systems by presenting mappings between binary parameters and accretion disk as well as ejecta masses, based on current fits to numerical relativity simulations (Sec.~\ref{sec:param_space}). We estimate that disk ejecta may dominate over dynamical ejecta across most of the parameter space for NSBH systems and BNS systems with a light-to-intermediate secondary neutron star. Dynamical ejecta may dominate in BNS systems with either very high mass ratio or a heavy secondary star.

We argue that to lowest order, the nucleosynthesis and kilonova signatures of post-merger accretion disks are determined by a single parameter, the (initial) accretion rate $\dot{M}$, which is largely determined by the initial mass of the disk. We demonstrate the existence of an ignition threshold $\dot{M}_{\rm ign}$ for weak interactions by employing a simplified 1D analytical disk model and self-consistent 3D GRMHD simulations with weak interactions. This threshold divides the 1D parameter space of post-merger accretion disks into two distinct regions (radiative efficient versus advection dominated) that lead to qualitatively different behavior with regard to nucleosynthesis and kilonova emission. We explore this change in behavior by investigating three disk models across this threshold in detail. Some key findings from these simulations include:

\begin{itemize}
	\item 
	We find that the radiative efficiency---the fraction of accreted rest mass energy turned into radiation per unit time---is very small below the ignition threshold ($\lesssim\!10^{-3}$). It rises to almost the theoretical maximum ($\gtrsim 0.1$ for the configurations considered here) just above the ignition threshold $\dot M_{\rm ign} \sim 1\times 10^{-3} M_\odot s^{-1}$, and finally decreases with increasing $\dot{M}$ (or initial disk mass) as a result of increasing optical depth in the accretion flow that starts to trap neutrinos (Fig.~\ref{fig:lum-mdot}).

	\item Protonization timescales $t_{Y_e}$ for the outer (non-degenerate) parts of our accretion flows vary strongly, with approximate scaling $t_{Y_e} \propto \dot M^{-5/4}$ (Figs.~\ref{fig:panels_Ye_eta} and \ref{fig:panels_R_eff_entropy}). This leads to massive accretion disks acquiring a high-$Y_{e}$ tail of ejecta much faster than lighter accretion disks (Fig.~\ref{fig:mass-Ye-outflow}). For the inner disks, we report a qualitative difference in compositional evolution across the ignition threshold. For high-mass disks above the threshold, high accretion rates cause dense accretion flows and force electrons to become degenerate. Weak interactions then lead to a self-regulation mechanism that keeps the disk at moderate degeneracy $\eta\sim1$ and high neutron richness $Y_e \approx 0.1$. For the low-mass disk below the ignition threshold studied here, negligible neutrino cooling causes a `pile-up' of heat in the inner disks, which triggers $e^+ e^-$ pair production. The injection of additional $e^+$ strongly accelerates protonization of the neutron-rich disk material in the vicinity of the black hole (Figs.~\ref{fig:panels_Ye_eta} and \ref{fig:panels_R_eff_entropy}).

	\item All three disk models investigated here produce large amounts of heavy $r$-process elements beyond the second r-process peak---corresponding to red kilonova components (cf.~Figs.~\ref{fig:abundances} and \ref{fig:mass_frac_peaks}). Our highest-$\dot{M}$ model is able to produce light $r$-process elements as well, with an overall r-process pattern in good agreement with solar abundances. This is the result of $(i)$ rapid protonization of the disks at high accretion rates and $(ii)$ self-irradiation of the outflows due to strong neutrino radiation. Our results indicate that disks close to or below the ignition threshold show heavily suppressed light r-process patterns. These disks largely keep their initial composition---weak interactions are suppressed to a level that protonization occurs on timescales longer than ejecta generation. Blue kilonova components are thus only expected for massive disks well above the ignition threshold.

	\item The effect of an ignition threshold is also imprinted on the bulk properties of the disk ejecta. We find that disks above the ignition threshold eject $\sim 30-50\%$ of their initial disk mass, whereas disks below the ignition threshold can eject $\gtrsim 60\%$ (estimates are to be taken with at least 10\% uncertainty). This increase is a result of the strong heating-cooling imbalance throughout the disk and its corona in absence of energetically significant cooling (Figs.~\ref{fig:panels_R_eff_entropy} and \ref{fig:entropy_space-time}). The increase of viscous heating with decreasing disk mass also enhances the high-velocity tail of the disk ejecta (Fig.~\ref{fig:outflows}).

	\item Outflows driven by MHD turbulence, with an effective $\alpha$ viscosity parameter and thus accretion rate set self-consistently (cf.~Sec.~\ref{sec:viscosity}), are generated early on during the disk evolution. We note that $\sim\!50\%$ of the total ejecta has already left the detection sphere at 1000\,km radius within the simulated time frame, a fraction of the effective viscous timescale of the disks (cf.~Tab.~\ref{tab:torusenergytab}, Sec.~\ref{sec:ejecta}). This is in contrast to $\alpha$-viscosity disks, in which ejecta is generated on much longer (viscous) timescales \citep{fernandez_long-term_2019,fernandez_landscape_2020,kyutoku_possibility_2020}. This has important consequences for ejecta composition and kilonova colors: for MHD disks there is less time for ejecta material to protonize and produce light r-process elements and blue kilonova components.

	\item \emph{Actinide-boost stars:} One immediate conclusion from our simulations is that disks below the ignition threshold retain their original composition (which may be arbitrarily neutron-rich, depending on the initial cold NS matter) and can thus contribute to \emph{actinide-boost events}---ejecta with an overabundance of actinides relative to lanthanides. Environments polluted by such ejecta could be conducive to forming actinide-boost stars. Roughly $\sim\!30\%$ of all r-process enhanced stars show such an overabundance \citep{Mashonkina2014}. This has so far been attributed to NS merger events with dominating and very neutron-rich tidal ejecta (e.g., \citealt{holmbeck_actinide_2019}). Here we demonstrate that, depending on the EOS, post-merger accretion disks below the ignition threshold could contribute to actinide overabundance to a similar or even higher level than dynamical ejecta (cf.~right panel of Fig.~\ref{fig:bns_parameters}, and considering that such disks may evaporate $>60\%$ of disk material into outflows). Above the ignition threshold, the lanthanide-heavy disk ejecta will tend to `wash out' possible actinide overabundance from dynamical ejecta, unless lanthanide production is severely suppressed at very high accretion rates \citep{miller_full_2019,li_neutrino_2021}. Furthermore, for NSBH systems, our parameter study suggests that actinide overabundance may mainly result from post-merger accretion disks, rather than dynamical ejecta (cf.~Fig.~\ref{fig:nsbh_parameters}). This is because disk ejecta may dominate over most of the parameter space. Actinide overabundance can thus be realized for disks below the ignition threshold, while above the ignition threshold potential actinide overabundance from dynamical ejecta may be (over)compensated by lanthanide-heavy ejecta from the disk (with the same caveat at high accretion rates mentioned for BNS systems above). This is also relevant to inferences about metal-poor star populations (e.g., \citealt{Holmbeck2020}).

\end{itemize}

Finally, we note a few caveats and limitations of the present study:

\begin{itemize}
	\item The late-time viscous regime of the outflows is not simulated here for reasons of computational cost. These late-time outflows may moderately enhance the high-$Y_{e}$ tail of the ejecta distribution \citep{fernandez_long-term_2019,fernandez_landscape_2020,Metzger:2008jt,metzger_time-dependent_2008}. However, given that the disks have already undergone significant viscous spreading (cf.~Fig.~\ref{fig:mdot-t}), we expect weak interactions to be close to freeze-out already, so that compositional changes may be moderate (in particular for disks below the ignition threshold).

	\item Additional variation in the total mass of MHD outflows on the order of $\sim\!10\%$ may result from variation of initial magnetic field configurations \citep{Christie:2019lim}, which we do not explore here, but rather defer to future work.

	\item At accretion rates higher than simulated here, more accurate neutrino transport is required to reliably capture the effect of neutrino self-irradiation \citep{miller_full_2019,li_neutrino_2021}.
\end{itemize}

The current suite of simulations represents a first step in exploring the parameter space of post-merger accretion disks in GRMHD. Future studies along the lines of \citet{fernandez_landscape_2020} are required to investigate the sensitivity of ejecta properties to black-hole mass and spin, initial compactness of the disk, initial magnetic field configurations, etc., to a level that is not yet encoded in the accretion rate parameter, and to probe a wider range in $\dot{M}$ systematically.

\acknowledgements
The authors thank Brian Metzger, Jonah Miller, Duncan Brown, Eric Coughlin, and Stefan Ballmer for discussions. This work was performed in part at the Kavli Institute for Theoretical Physics; SD and DMS thank the organizers and participants of the KITP program ``The New Era of Gravitational-Wave Physics and Astrophysics'' for stimulating discussions and KITP for hospitality. This research was supported in part by the National Science Foundation under Grant No. NSF PHY-1748958. Resources supporting this work were provided by the NASA High-End Computing (HEC) Program through the NASA Advanced Supercomputing (NAS) Division at Ames Research Center and by Syracuse University. This research was enabled inpart by support provided by SciNet (www.scinethpc.ca) and Compute Canada (www.computecanada.ca). SD acknowledges support from the Research Excellence Doctoral Fellowship at Syracuse University and the Director's Postdoctoral Fellowship at Los Alamos National Laboratory---with the Laboratory Directed Research and Development Program project number 20200687PRD2. Los Alamos National Laboratory is operated by Triad National Security, LLC, for the National Nuclear Security Administration of U.S. Department of Energy under Contract No. 89233218CNA000001. DMS acknowledges the support of the Natural Sciences and Engineering Research Council of Canada (NSERC), funding reference number RGPIN-2019-04684. Research at Perimeter Institute is supported in part by the Government of Canada through the Department of Innovation, Science and Economic Development Canada and by the Province of Ontario through the Ministry of Colleges and Universities.

\appendix

\section{Ignition threshold and simplified 1D disk model}
\label{app:ignition_threshold}

For completeness, we provide the expression for $F(x,\chi_{\rm BH})$ that appears in the simplified one-dimensional $\alpha$-disk model described in Sec.~\ref{sec:ignition_threshold}:
\begin{eqnarray}
	F(x,\chi_{\rm BH}) &\equiv& \frac{x^3 + \chi_{\rm BH}}{(x^3 - 3x + 2\chi_{\rm BH})^{1/2}x^{3/2}} \Bigg[ (x-x_0)-\frac{3}{2}\chi_{\rm BH} \ln\frac{x}{x_0}-\frac{3(x_1-\chi_{\rm BH})^2}{x_1(x_1-x_2)(x_1-x_3)}\ln\left(\frac{x-x_1}{x_0-x_1}\right) \nonumber\\ 
	&&- \frac{3(x_2-\chi_{\rm BH})^2}{x_2(x_2-x_1)(x_2-x_3)}\ln\left(\frac{x-x_2}{x_0-x_2}\right) -\frac{3(x_3-\chi_{\rm BH})^2}{x_3(x_3-x_1)(x_3-x_2)}\ln\left(\frac{x-x_3}{x_0-x_3}\right)\Bigg]. \label{eq:F_x}
\end{eqnarray}
Here, $x\equiv \sqrt{2r/r_{\rm g}}$, $x_0$ corresponds to the location of the marginally stable orbit, and $x_1$, $x_2$, $x_3$ are the roots of $x^3-3x+2\chi_{\rm BH}=0$. Explicitly, $x_1=2\cos(\frac{1}{3}\cos^{-1}\chi_{\rm BH}-\pi/3)$, $x_2=2\cos(\frac{1}{3}\cos^{-1}\chi_{\rm BH}+\pi/3)$, and $x_3=-2\cos(\frac{1}{3}\cos^{-1}\chi_{\rm BH})$. See also \citet{page_disk-accretion_1974} and \cite{
chen_neutrino-cooled_2007}.



\end{document}